\documentclass[twocolumn,twocolappendix]{aastex631}

\newcommand{\rh}[1]{#1}

\usepackage{amsmath}
\usepackage{physics}
\usepackage{bbold}
\usepackage{soul}
\received{March 11, 2024}
\revised{July 18, 2024}
\accepted{July 19, 2024}

\submitjournal{ApJ}

\graphicspath{{./}{FinalFigures/}}

\begin{document}

\title{Cosmic Ray Feedback on Bi-stable ISM Turbulence}


\correspondingauthor{Roark Habegger}
\email{rhabegger@wisc.edu}

\author[0000-0003-4776-940X]{Roark Habegger}
\affiliation{Astronomy Department, University of Wisconsin-Madison}

\author[0000-0003-3328-6300]{Ka Wai Ho}
\affiliation{Astronomy Department, University of Wisconsin-Madison}
\affiliation{Theoretical Division, Los Alamos National Laboratory}

\author[0000-0003-1683-9153]{Ka Ho Yuen}
\affiliation{Theoretical Division, Los Alamos National Laboratory}

\author[0000-0003-4821-713X]{Ellen G. Zweibel}
\affiliation{Astronomy Department, University of Wisconsin-Madison}
\affiliation{Physics Department, University of Wisconsin-Madison}

\begin{abstract}%
While cosmic rays $(E\gtrsim 1\,\mathrm{GeV})$ are well coupled to a galaxy's interstellar medium (ISM) at scales of $L>100\,\mathrm{pc}$, adjusting stratification and driving outflows, their impact on small scales is less clear. Based on calculations of the cosmic ray diffusion coefficient from observations of the grammage in the Milky Way, cosmic rays have little time to dynamically impact the ISM on those small scales. Using numerical simulations, we explore how more complex cosmic ray transport could allow cosmic rays to couple to the ISM on small scales. We create a two-zone model of cosmic ray transport, with the cosmic ray diffusion coefficient set at the estimated Milky Way value in cold gas but smaller in warm gas. We compare this model to simulations with a constant diffusion coefficient. 
Quicker diffusion through cold gas allows more cold gas to form compared to a simulation with a constant, small diffusion coefficient. However, slower diffusion in warm gas allows cosmic rays to take energy from the turbulent cascade anisotropically. This cosmic ray energization comes at the expense of turbulent energy which would otherwise be lost during radiative cooling. Finally, we show our two-zone model is capable of matching observational estimates of the grammage for some transport paths through the simulation.

\end{abstract}


\section{Introduction} \label{sec:intro}

Cosmic rays, along with magnetic fields, turbulent energy, thermal energy, and radiation pressure, are a major constituent of the interstellar medium (ISM, see \citealt{MO07}). Cosmic ray pressure, provided primarily by protons of a few GeV, contributes to the vertical stratification of galactic disks, drives galactic winds, impedes star formation, and alters the evolution of supernova remnants (see \citealt{2017Zweibel}, \citealt{2023Ruszkowski}, and \citealt{2023Owen} for recent reviews and references). 

One of the major unresolved issues in cosmic ray astrophysics is how the cosmic rays derive their energy. The favored and dominant mechanism is acceleration in shocks driven by supernovae, with a complementary role for shocks in stellar winds \citep{1977Axford,1978Bell1,1978Blandford}. But the extent to which cosmic rays tap other sources of energy (e.g. energy in large scale turbulence \citep{1988Ptuskin,2022Bustard}) throughout their lifetime is still unclear.

A mechanism for converting turbulent energy to cosmic ray energy was suggested by \cite{1988Ptuskin} and has recently been revived by \cite{2019Commercon}, \cite{2019Dubois}, \cite{2022Bustard}, \cite{2023Bustard}. This mechanism works as follows: cosmic rays are scattered by kinetic scale fluctuations in the ambient magnetic field (a standard feature of cosmic ray transport theory since the foundational work of \citealt{1965Parker} \& \citealt{1966Jokipii}). On scales larger than the mean free path between scatterings the cosmic rays behave as a diffusive fluid. This diffusion allows cosmic rays to extract energy from compressive fluctuations much as a thermally conducting gas absorbs heat from sound waves. The energy gain is maximized when the diffusion time across a turbulent eddy is comparable to the eddy turnover time.  

The value of the cosmic ray diffusion coefficient in the Milky Way disk is derived mainly from the observed abundances of cosmic rays produced by spallation reactions between cosmic rays and interstellar gas. Those abundances are essentially a column density measurement over the convoluted paths cosmic rays follow. Recent measurements estimate the diffusion coefficient to be $\kappa_\parallel \sim 10^{28 } \, \mathrm{cm}^2 \, \mathrm{s}^{-1}$ \citep{2019Evoli,2020Evoli}. Taking a reasonable driving scale for ISM turbulence of $100\,\mathrm{pc}$ and a realistic phase velocity of $10\,\mathrm{km}\,\mathrm{s}^{-1}$ gives a characteristic eddy turnover time of $\tau \sim 10 \,\mathrm{Myr}$. On the same length scale, the cosmic rays will diffuse in just $\tau \sim 0.3 \,\mathrm{Myr}$. This fundamental mismatch suggests that turbulent energy extraction and feedback on ISM scales by cosmic rays is weak. This conclusion is corroborated by \cite{2019Commercon,2022Bustard,2023Bustard}. Achieving maximum coupling would require a diffusivity of order $\sim 10^{26} \, \mathrm{cm}^2 \, \mathrm{s}^{-1}$.

The effects of cosmic rays on turbulence are not limited to extraction of energy. In a bi-stable medium, decreasing the cosmic ray diffusion coefficient to a small value will limit the production of cold gas \citep{2019Commercon}. This result comes from cosmic rays being trapped in condensing clouds instead of diffusing out. While stuck in the clouds, they provide a significant pressure against condensation and collapse. When cosmic ray energy gain is maximized (when the eddy turnover time and cosmic ray diffusion time scale are similar), there are some  circumstances under which the turbulent energy cascade is truncated at $L\lesssim 10$ pc scales \citep{2023Bustard}. Both of these effects would be in conflict with observations, which show plenty of cold gas formation and no cutoff in the turbulent energy cascade down to a scale of $L \lesssim 1\,\mathrm{pc}$ \citep{2018Pingel,2022Pingel}.

These past works on cosmic ray coupling to turbulence generally assume a constant diffusion coefficient. To test if the conclusions remain true for more complex cosmic ray transport, we examine the case where $\kappa_\parallel$ is not a spatial constant. This idea is similar to two-zone and multi-zone models used in studies of cosmic ray propagation \citep{2016Guo,2019Johannesson,2023DeLaTorreLuque}, and has a sound physical basis: in weakly ionized gas, the small scale magnetic field fluctuations which cosmic rays scatter off are strongly damped by friction between ions and neutrals (ion-neutral damping, see \citealt{1971Kulsrud}). Therefore, it is plausible that $\kappa_{\parallel}$ is larger in the denser, more neutral phases of the ISM. 

Instead of tracking or evolving ionization in our simulations, we use temperature dependence as a proxy. We have the cosmic rays diffuse slowly through warm gas and quickly through cold gas as a result of ion-neutral damping. This model is derived from a study of cosmic ray driven galactic outflows and star formation feedback in global galaxy simulations by \cite{2018Farber}. In that paper, the cosmic ray diffusion coefficient was assumed to change between hot and warm gas, resulting in lower density winds and more star formation.

Our goal with implementing a variable diffusion coefficient and applying it to a bi-stable ISM is to reconsider some of the results from \cite{2019Commercon,2022Bustard,2023Bustard}. \textbf{More specifically, can we form cold gas while still having cosmic rays significantly impact the properties of the turbulent, multiphase ISM?}

In this paper, we demonstrate that the two zone model allows cosmic rays to extract significant energy from turbulence in warm, low density gas while reducing their dynamical role in cold gas.  Additionally, the cosmic rays accumulate the column density required to reproduce observed values of the grammage. This result suggests that observations of light element abundances might not imply cosmic rays are dynamically insignificant on small scales $(L\lesssim 100 \,\mathrm{pc})$. 

Up to now, we have not addressed the source of the fluctuations that scatter cosmic rays, leading to a diffusive random walk through the ISM. They could be generated by the cosmic rays themselves, through the streaming instability \citep{1969Kulsrud} or simply be part of an extrinisic turbulent cascade. There are significant differences: streaming transport introduces an extra heating term which adjusts how cosmic rays feed back on the thermal gas \citep{2013Zweibel,2017Zweibel}. Additionally,
opposing cosmic ray pressure and plasma density gradients can result in cosmic ray pressure plateaus, or bottlenecks \citep{2017Wiener,2021Bustard}. However, to keep the problem simpler and better connect with past work, we neglect the streaming instability in this paper. For the impact of streaming on cosmic ray feedback in the turbulent ISM, we direct readers to \cite{2019Dubois} and \cite{2023Bustard}.

We compare and contrast four 3D turbulent box $(L\sim 100\,\mathrm{pc})$ simulations: one with no cosmic rays, one with constant diffusivity set to the canonical Milky Way value $\kappa_\mathrm{MW}=3\times 10^{28}\mathrm{cm}^2\mathrm{s}^{-1}$ (note this is slightly larger than recent observations suggest), another with constant diffusivity set to the critical value for maximum energy gain found in previous studies $(\sim 10^{26}\mathrm{cm}^2\mathrm{s}^{-1})$, and one which uses a temperature dependent cosmic ray energy diffusion coefficient which switches between the two values. Other than cosmic ray energy and transport, the simulations have identical initial conditions. They even use the same random seed for turbulent driving. Using identical setups implies the differences we observe are the result of including cosmic ray energy and varying its transport. We run our simulations for more eddy turnover times than previous works, adding to the statistical robustness of our analysis. While the exact diffusion coefficients in some of our simulations may not apply to the Milky Way galaxy's ISM, our work improves the physical understanding of cosmic ray feedback on a turbulent cascade. 

This paper is structured as follows: in Section \ref{sec:methods}, we detail our computational methods and choice of simulation parameters; in Section \ref{sec:results} we show results from the simulations; in Section \ref{sec:disc} we discuss our results, along with their significance; and in Section \ref{sec:conclusions} we provide conclusions and a short summary of the work. In Appendix \ref{sec:appendix:LowRes} we provide some additional results, from lower resolution simulations, which add to the robustness of our conclusions.

\section{Methods} \label{sec:methods}
In this section, we detail the methods used for cosmic ray hydrodynamics in MHD, discuss the limits of this implementation of cosmic ray hydrodynamics, our use of turbulent driving, the implementation of radiative cooling, the addition of a scalar dye for tracking cold gas formation, and our method for cosmic ray decoupling in cold gas. Finally, we derive and discuss our simulation parameters. 

We use a recent release of the \texttt{Athena++} magnetohydrodynamic (MHD) code which evolves total cosmic ray energy and flux alongside the other hydrodynamic variables \citep{2018Jiang,2020Stone}. We start the simulation with a 3D homogeneous box of equal sides in each direction $(L=100\,\mathrm{pc})$ with periodic boundary conditions. The free parameters of the initial setup are the temperature $T$ and density $\rho$ of the thermal gas, the  plasma beta $\beta \equiv 8\pi P_g / B^2$ (where $P_g$ is the thermal pressure and $B$ is the magnetic field strength), and the cosmic ray beta $\beta_\mathrm{cr} \equiv P_g/P_c$ (where $P_c$ is the cosmic ray pressure). The magnetic field is initially uniform and directed in the along $\hat{x}$. The simulations presented here have a resolution $\Delta x \sim 0.39 \,\mathrm{pc}$, but we include results from lower resolution simulations $(\Delta x \sim 0.78 \,\mathrm{pc})$ in Appendix \ref{sec:appendix:LowRes}.


\subsection{MHD \& CR Hydrodynamics} \label{sec:methods:CRMHD}
The \texttt{Athena++} implementation from \cite{2018Jiang} solves the following equations:

\begin{equation}
    \pdv{\rho}{t} + \div{ \left( \rho \vb{u} \right)} = 0
    \label{eqn:cont}
\end{equation}

\begin{equation}
    \pdv{\vb{B}}{t} - \curl{\left( \vb{u} \times \vb{B} \right)} = 0
    \label{eqn:induct}
\end{equation}

\begin{eqnarray}
    \pdv{\rho \vb{u}}{t} + \div{ \left( \rho \vb{u}\vb{u} + \mathbb{1}\left(P_g + \frac{B^2}{2}\right) - \vb{B} \vb{B} \right)} =  \nonumber\\ 
    \sigma_c \cdot \left(\vb{F}_c - \frac{4}{3}E_c\vb{u} \right)
    \label{eqn:mom}
\end{eqnarray}

\begin{eqnarray}
    \pdv{E}{t} + \div{ \left( \left( E+ P_g + \frac{B^2}{2} \right)\vb{u}  - \left(\vb{B} \cdot \vb{u}\right)\vb{B} \right)} =  \nonumber\\ 
    \mathcal{L}(n,T)-\frac{1}{3}\vb{u} \cdot \grad{E_c}
    \label{eqn:e}
\end{eqnarray}

\begin{equation}
    \pdv{E_c}{t} + \div{ \vb{F}_c} = \frac{1}{3}\vb{u} \cdot \grad{E_c}
    \label{eqn:cre}
\end{equation}

\begin{equation}
    \frac{1}{V_m^2}\pdv{\vb{F}_c}{t} + \frac{1}{3}\grad{ E_c} = - \sigma_c \cdot \left(\vb{F}_c - \frac{4}{3}E_c\vb{u} \right).
    \label{eqn:crflux}
\end{equation}

In the above equations, the variables evolved in time are the thermal gas density $\rho$, bulk velocity $\vb{u}$, gas pressure $P_g$, magnetic field $\vb{B}$, cosmic ray energy $E_c$, and cosmic ray energy flux $\vb{F}_c$. The total magnetohydrodynamic energy density (excluding $E_c$) is a combination of those variables: $E \equiv \rho u^2/2 + P_g/(\gamma_g - 1) + B^2/2$. Throughout this work the thermal gas is treated as ideal with ratio of specific heats $\gamma_g = 5/3$ and the cosmic rays are treated as an ultrarelativistic fluid with $\gamma_c = 4/3$. The radiative heating and cooling function $\mathcal{L}(n,T)$ depends on number density $n = \rho/\overline{m}$ (where $\overline{m}$ is the mean particle mass) and gas temperature $T = P_g/(nk_B)$ (where $k_B$ is the Boltzmann constant). This function is specified in Section \ref{sec:methods:cooling}.

For the evolution of cosmic ray energy, there are two other key variables. First, the modified speed of light $V_m$ appears where the speed of light should be. For numerical accuracy, the only requirement on $V_m$ is it needs to be the fastest speed in the simulation, which is essentially requiring:
\begin{equation}
    V_m^2 > u^2 +  \frac{P_g + P_c + \frac{B^2}{2}}{\rho}.
\end{equation}
Second, we define the cosmic ray transport matrix $\sigma_c$ to allow a split of parallel and perpendicular diffusion of cosmic rays:
\begin{equation}
    \sigma_c = \frac{1}{\kappa_\perp} \mathbb{1} + \left(\frac{1}{\kappa_\parallel} - \frac{1}{\kappa_\perp} \right)\hat{b}\hat{b}.
    \label{eqn:transport}
\end{equation}
The magnetic field direction $\hat{b}$ is calculated locally, in each cell at each time, and is not in general the mean magnetic field direction. We set $\kappa_\perp$ to be a small value, $10^{18}\,\mathrm{cm}^2 \,\mathrm{s}^{-1}$, so that perpendicular diffusion across a cell width will take $10\times$ the simulation run time. This choice of perpendicular diffusion coefficient is justified by modern cosmic ray transport models. Quasilinear diffusion theory predicts $\kappa_{\perp}/\kappa_{\parallel} \sim (r_g/l)^2 \sim 10^{-11}$ for gyroradius $r_g \sim 1 \,\mathrm{AU}$ and parallel mean free path $l \sim 1 \, \mathrm{pc}$  \citep{2014Desiati}. From another perspective, we could account for perpendicular transport due to unresolved fluctuations (below $\sim 1\,\mathrm{pc}$) in the magnetic field. This method still finds a decreased perpendicular diffusion coefficient ($\kappa_\perp/\kappa_\parallel\propto M_A^n \lesssim (0.5)^n$ for $n=2-4$ depending on models, see \citealt{2003Matthaeus,2006Shalchi,YanL08}). Because we use an extremely low value of $\kappa_{\perp}$, cosmic ray energy barely diffuses perpendicular to the field lines over the course of our simulation. We allow a more complex definition for the parallel transport $\kappa_\parallel$, which is detailed in Section \ref{sec:methods:decouple}.

\subsection{Limits of the Fluid Prescription for Cosmic Rays} \label{sec:methods:FluidLimit}
Since we use a fluid description of cosmic rays rather than a full kinetic description or evolution of particle motion, it would be counterproductive to resolve physical scales shorter than the mean free path of scatterings $(l\sim\kappa_{\parallel}/c)$. For the largest $\kappa_{\parallel}$ we consider, the length scale is $l\sim 0.32 \,\mathrm{pc}$. The resolution of our simulation is $\Delta x = 0.39 \,\mathrm{pc} \, L_{100}/ N_{256}$ where $N_{256} = N/256$ is the number of cells in each direction in terms of our base simulations and $L_{100} = L/(100 \mathrm{pc})$ is the simulation box length. The resolution we use is $256^3$; higher resolution would not be meaningful when using the fluid treatment of cosmic rays. We also include $128^3$ resolution results in Appendix \ref{sec:appendix:LowRes}.

Going beyond a fluid description, e.g. by running MHD-PIC (Particle In Cell, e.g. \citealt{Athena_MHDPIC}) simulations, which use PIC for cosmic ray transport, could eliminate this restriction on resolution and would self-consistently evolve the cosmic ray distribution function. But that would further limit the time step beyond what we manage and require an exorbitant amount of computational resources to match our simulations' size. 

\subsection{Turbulent Driving}\label{sec:methods:driving}

We drive our initially static system at large scales by injecting energy at wavenumbers $m=1/L=k/(2\pi)=\{1,2\}$ (equivalent to $L=100,50\,\mathrm{pc}$ in physical units), with a $k^{-2}$ dependence. We use a constant total energy injection rate 
\begin{equation}
    \dot{E} = \frac{E_\mathrm{th}}{\tau} = 5.5 \cdot 10^{48} \frac{\mathrm{erg}}{\mathrm{Myr}}  \left( \frac{11 \, \mathrm{Myr}}{\tau}\right) T_4 n_0 L_{100}^3
    \label{eq:driving}
\end{equation}
where $T_4 = T/ (10^4\, \mathrm{K})$, $n_0 = n / (1\mathrm{cm}^{-3})$, and $\tau$ is the characteristic time to inject energy equal to the initial thermal energy in the simulation volume. 

The \texttt{Athena++} turbulent driving module uses an Ornstein-Uhlenbeck process \citep{1930Uhlenbeck} with a compressive-solenoidal splitting \citep{1988Eswaran}. In our case, the energy is injected as velocity perturbations using purely compressive driving, with a forcing function $\vb{F}$ which is derived from the velocity perturbations under the assumption that the energy input rate, $\vb{u} \cdot \vb{F}$, is constant. 

The forcing can be decaying (only injected at the start of the simulation), continuous (driven at every time step with a user defined time decorrelation), or impulsive (driven at a time interval set by the user). We use impulsive driving because of the small time step imposed by the modified speed of light $V_m$ in Equation \ref{eqn:crflux}. The impulses are driven at a timestep of $\delta t = 0.005$ in computational units. This timestep is near the value of the timestep necessary to integrate fluid motion without the cosmic ray energy equations. Therefore we are doing the equivalent of continuous driving in a simulation without cosmic rays.

\subsection{Radiative Cooling} \label{sec:methods:cooling}

We adopt the simple model of optically thin radiative cooling from \cite{2006Inoue} which accounts for the emission of Ly$\alpha$ photons (important in warm gas) and the emission of $[\mathrm{CII}]$ photons (important in cold gas). We assume heating as a result of the photoelectric effect on dust grains. These assumptions give the following radiative heating and cooling function:
\begin{equation}
    \mathcal{L}(n,T) = n \Gamma - n^2 \Lambda(T)
    \label{eq:radiation}
\end{equation}
where the cooling function is
\begin{eqnarray}
    \Lambda(T) =  \Bigg( 7.3\cdot 10^{-21} \exp\left[ \frac{-118400}{T+1500}\right] \nonumber &\\
    + 7.9\cdot 10^{-27} \exp\left[ \frac{-92}{T}\right] \Bigg) &
    \frac{\mathrm{erg}\,\mathrm{cm}^{3}}{\mathrm{s}}
    \label{eq:cooling}
\end{eqnarray}
and the heating rate is
\begin{equation}
    \Gamma = 2\cdot 10^{-26} \frac{\mathrm{erg}}{\mathrm{s}}.
    \label{eq:heating}
\end{equation}

\begin{figure}
    \centering
    \includegraphics[width=0.48\textwidth]{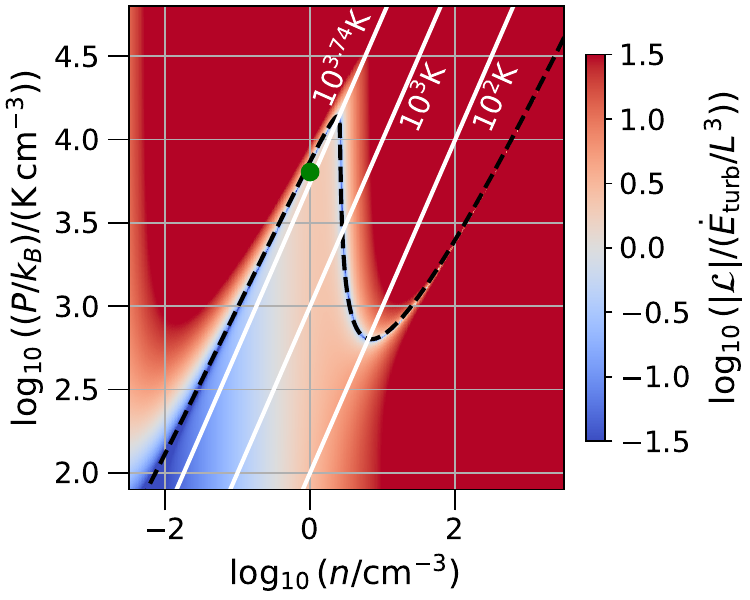}
    \caption{Comparison of energy sources and sinks in thermal gas phase space. Dashed black line shows where the gas is in thermal equilibrium ($\mathcal{L}=0$ in Equation \ref{eq:radiation}). Dashed white lines show constant temperature contours for $T=10^{3.74}\mathrm{K} = 5.5 \cdot 10^3 \mathrm{K}$, $T=10^{3}\mathrm{K}$, and $T=10^{2}\mathrm{K}$. Color map compares net radiative energy change (either loss via cooling or gain via heating) to the total turbulent energy we inject (Equation \ref{eq:driving}). The green dot shows where the simulations start, slightly below the equilibrium curve. Three phases (a warm phase, an unstable phase, and a cold phase) with different temperatures can exist in pressure equilibrium for pressures between $P/k_B \sim 10^3 \,\mathrm{K}\,\mathrm{cm}^{-3}$ and $P/k_B \sim 10^4 \,\mathrm{K}\,\mathrm{cm}^{-3}$. }   
    \label{fig:EradEturb}
\end{figure}

The equilibrium curve in pressure-density phase space created by these heating and cooling terms is shown as a black dashed line in Figure \ref{fig:EradEturb}. Using this radiative heating and cooling setup creates a gas which has two thermally stable phases (with different temperatures) which can exist in pressure equilibrium with each other. There is also a thermally unstable phase in the same pressure range. This bi-stable model of the ISM captures the important physical process of thermal instability \citep{1965Field} in its simplest form: the transition between a warm phase and a cold phase. 

Figure \ref{fig:EradEturb} also shows a colormap comparing the net heating and cooling rate with the turbulent energy density injection rate. The radiative energy rate is calculated by taking the absolute value of $\mathcal{L}(n,T)$ in Equation \ref{eq:cooling}, whereas the turbulent energy density injection rate is a constant in $(P,n)$ space (take Equation \ref{eq:driving}, and divide by simulation size $L^3=(100\,\mathrm{pc})^3$). While near thermal equilibrium (the black dashed line) the turbulent energy injection dominates, its contribution is subdominant in most of the domain (except for the blue wedge at low densities and low pressures). Note that the turbulent energy input is given for comparison purposes only; it is not a heating mechanism in the same sense as the photoelectric effect on dust grains. When the only energy gains and losses are those represented by Equation \ref{eq:radiation}, heating dominates below the thermal equilibrium curve and cooling dominates above it.

We start the simulation in thermally stable equilibrium with the gas in the warm phase at a temperature of $6388\,\mathrm{K}$ and a density of $1\,\mathrm{cm}^{-3}$. The turbulent stirring drives compression and rarefaction, with some gas getting pushed into thermally unstable states. A portion of the unstable gas then collapses into small scale, high density clouds via the thermal instability.

\subsection{Scalar Dye for Dense Gas Formation} \label{sec:methods:dye}

Although we do not include self gravity or mass sinks representing star particles in our simulations, we do attempt to track the formation of cold gas which could eventually form stars. We do this by injecting a passive scalar dye at a constant rate of $1 M_\odot \,\mathrm{Myr}^{-1}$ into cells which hit the temperature floor $(T_\mathrm{floor} = 20 \,\mathrm{K})$. This rate comes from assuming a star formation rate of $1 M_\odot \,\mathrm{yr}^{-1}$ in a galaxy volume of about $10^{12}\,\mathrm{pc}^3$, but then adjusting for our simulation volume of $10^{6}\,\mathrm{pc}^3$. 

The gas density in thermal equilibrium at the temperature floor $T=20\,\mathrm{K}$ is $252\,\mathrm{cm}^{-3}$. For our $N=256^3$ simulations, the gas mass in a single cell at that density and temperature is about $3.5 M_{\odot}$. The Jeans mass at that density and temperature is about $50 M_{\odot}$. Therefore, we would need to produce a contiguous agglomeration of $\sim 15$ cells in order to meet the conditions for gravitational instability.

Another consideration is self-shielding by molecular hydrogen in dense clouds. This effect changes the cooling function and allows more molecular gas to form, which could make it easier for a cloud to gravitationally collapse. Self-shielding becomes important at an atomic hydrogen column density of $N_H \sim 5 \cdot 10^{20} \,\mathrm{cm}^{-2}$ \citep{Draine2011}, which corresponds to a number density 
\begin{equation}
    n_H \sim \frac{N_H}{R} = 162 \,\mathrm{cm}^{-3} \frac{N_{H}}{5 \cdot 10^{20}\,\mathrm{cm}^{-2}} \frac{R}{1\,\mathrm{pc}}
\end{equation}
assuming a cloud of radius $R=1\,\mathrm{pc}$. This effect may therefore become important in the most central regions of any cold gas regions in our simulations. However, we certainly do not resolve these molecular regions, given our resolution of $\Delta x = 0.39\,\mathrm{pc}$ in the $256^3$ simulations. We do not include their formation or the effect hydrogen self shielding because of this resolution limit. 

While we do not simulate star formation, the effects of self gravity, or the formation of molecular gas, the scalar dye traces dense gas formation over the simulation's evolution. Overall, it allows us to track how much total dense gas is formed in each simulation. 

\subsection{Two Zone Cosmic Ray Diffusion Coefficient} \label{sec:methods:decouple}

Our two-zone cosmic ray transport model is motivated by ion-neutral damping \citep{2016ApJ...826..166X}. We assume the cosmic rays are scattered by Alfven waves from an extrinsic turbulent cascade. When the ionization of the ISM decreases in cold clouds, those Alfven waves are damped by collisions between ions and neutrals \citep{1969Kulsrud,1971Kulsrud}, increasing the effective cosmic ray diffusion coefficient in that gas. As we mentioned in Section \ref{sec:intro}, the effect of an increased diffusion coefficient due to ion-neutral damping in cold gas has been already examined in the context of global simulations of cosmic ray feedback \citep{2018Farber}. 

To implement the temperature dependent cosmic ray energy diffusion coefficient, we adjust the method used in  \cite{2018Farber}, which used a discontinuous, piecewise constant function for $\kappa_\parallel$. This prescription produces numerical instability if the discontinuity is too large. To increase numerical stability, we introduce a smooth function for the transition between two values of $\kappa_\parallel$ in the cold and warm gas; $\kappa_c$ and $\kappa_w$, respectively.
To allow larger variations in $\kappa$ we build the transition in logarithmic space
\begin{eqnarray}
    \log_{10} \kappa_\parallel(T) = \left( \log_{10} (\kappa_c) - \log_{10} (\kappa_w) \right)f_c(T) \nonumber\\ 
    + \log_{10} (\kappa_w)
    \label{eq:decouple}
\end{eqnarray}
using a switching function
\begin{equation}
    f_c(T) = \frac{1}{2}\left( 1 - \tanh \left(\frac{T-T_0}{\Delta T} \right) \right).
    \label{eq:switch}
\end{equation}
This function requires a transition temperature $T_0$ between the diffusion coefficient values, along with a transition width $\Delta T$ in temperature. Our implementation allows the diffusion coefficient to change by several orders of magnitude with more numerical stability.

We use $10^3\mathrm{K}$ for transition temperature $T_0$ because it cuts through the middle of the unstable neutral medium in the pressure-density phase space (see the $T=10^3\mathrm{K}$ contour in Figure \ref{fig:EradEturb}). We use a width of $\Delta T = 0.1 T_0$ which is wide enough to allow for smooth transitions and steep enough to limit the number of cells with a diffusion coefficient in between the two constant values.

\subsection{Parameter Choices} \label{sec:methods:params}

After considering all these different physical processes, we still have several parameters to set for the simulations. First, we set the gas density, temperature, plasma beta, and cosmic ray beta which describe the initially homogeneous fluid. Then we discuss possible values of the cosmic ray diffusion coefficient. Finally, we set the diffusion coefficient value in each simulation, including its value in the warm and cold gas ($\kappa_w$ and $\kappa_c$ in Equation \ref{eq:decouple}) for the simulation where we use our two-zone model.

The parameters for each simulation are shown in Table \ref{tab:sims}. We set the initial plasma parameters as $T_i = 6388 \, \mathrm{K}$, $n = 1 \mathrm{cm}^{-1}$, $\beta = 1$, and $\beta_\mathrm{cr} = 5$ (in the simulation without cosmic rays, $\beta_\mathrm{cr} = \infty$). The choice of $\beta_\mathrm{cr} = 5$ comes from the assumption that the cosmic ray energy increases with time and the radiative cooling causes thermal energy to decrease with time. Therefore, the $\beta_\mathrm{cr}$ should decrease with time. If we want there to be a significant amount of time where the thermal, magnetic, and cosmic ray energy are all a similar order of magnitude (i.e. $\beta_\mathrm{cr}\sim 1$), then we have to start with a larger $\beta_\mathrm{cr} > 1$. An approximate energy equipartition mimics the state of the ISM in the Milky Way.

\begin{table}[t]
    \centering
    \begin{tabular}{|l|r|r|r|r|r|} \tableline
         Simulation& 
         $T_i (\mathrm{K})$ & $\beta$ & $\beta_\mathrm{cr}$ &
         $\kappa_{w} (\mathrm{cm}^2 \, \mathrm{s}^{-1})$ & $\kappa_c (\mathrm{cm}^2 \, \mathrm{s}^{-1})$  \\\tableline
         \texttt{No CR} & 
         $6388$ & $1$ & $\infty$ &
         - & - \\\tableline 
         \texttt{Milky Way} & 
         $6388$ & $1$ & $5$ &
         $3\cdot 10^{28}$ & $3\cdot 10^{28}$ \\\tableline 
         \texttt{Critical} & 
         $6388$ & $1$ & $5$ &
         $5.82 \cdot 10^{25}$ &  $5.82 \cdot 10^{25}$ \\\tableline 
         \texttt{Two Zone} & 
         $6388$ & $1$ & $5$ &
         $5.82 \cdot 10^{25}$ & $3\cdot 10^{28}$ \\\tableline 
    \end{tabular}
    \caption{Start-up parameters and cosmic ray energy diffusion coefficients for each simulation. Each of these setups were run at two resolutions $(128,256)$, with a modified speed of light $V_m = (1/60, 1/30) c$ respectively. See Appendix \ref{sec:appendix:LowRes} for analysis of the $N=128$ simulations.}
    \label{tab:sims}
\end{table}

For the diffusion coefficients, we start by considering the average value in the Milky Way. The canonical Milky Way cosmic ray diffusion coefficient value of $\kappa_\mathrm{MW} \approx 3 \times 10^{28} \, \mathrm{cm}^2 \, \mathrm{s}^{-1}$ comes from older measurements of the grammage $X$, which has dimensions of mass per area \citep{1990Berezinskii,2001Jones}. Grammage is a measure of the amount of material a cosmic ray passes through on its path through the ISM and is calculated by the observed ratio of primary and secondary cosmic rays. The average diffusion coefficient $\overline{\kappa}_\parallel$ can be calculated from the grammage using
\begin{equation}
    X \approx \overline{\rho} c \tau = \overline{\rho} c \frac{H^2}{\overline{\kappa}_\parallel} \hspace{0.1in}
    \rightarrow \hspace{0.1in}
    \overline{\kappa}_\parallel \approx \overline{\rho} c \frac{H^2}{X }.
    \label{eqn:grammage}
\end{equation}
where $H$ is the root-mean-square displacement of cosmic rays from their source during their confinement time in the Milky Way, $\overline{\rho}$ is the average density of the ISM, and $\tau$ is the average time it takes the cosmic rays to diffuse. Taking displacement to be the half-thickness of the Milky Way $H \sim 250 \,\mathrm{pc}$, mean density to be $\overline{\rho} = m_p \,\mathrm{cm}^{-3}$, and $X \sim 1 \,\mathrm{g}\,\mathrm{cm}^{-2}$ for the low energy $(\sim 1\,\mathrm{GeV})$ cosmic rays, gives an order of magnitude estimate of $\overline{\kappa}_\parallel \sim 3 \cdot 10^{28} \, \mathrm{cm}^2 \, \mathrm{s}^{-1}= \kappa_\mathrm{MW}$. We stick with this older, canonical, estimate \citep{2001Jones} for our simulations. However, more modern estimates suggest a slightly larger grammage $X \sim 3 \,\mathrm{g}\,\mathrm{cm}^{-2}$ for the lowest energy cosmic rays, reducing the average diffusion coefficient by the same factor.  

Another interesting value is the critical diffusion identified in \cite{2019Commercon,2022Bustard,2023Bustard}. Near the critical diffusion coefficient the cosmic rays will significantly disrupt the turbulent cascade. \cite{2022Bustard} use analytical arguments to determine the dependence of the critical rate on various parameters, finding 
\begin{eqnarray}
    \kappa_\mathrm{crit} = 0.15 v_\mathrm{ph} L_0 
    = 0.15 L_0 \sqrt{\frac{P_g + P_B + P_c}{\rho}} \nonumber \\
    = 0.15 L_0 \sqrt{\frac{k_B T}{m}}\sqrt{1 + \beta^{-1} + \beta^{-1}_\mathrm{cr}} \nonumber \\
    = 5.82 \cdot 10^{25} \frac{\mathrm{cm}^2}{\mathrm{s}} \frac{L_0}{100 \mathrm{pc}} T_4^{1/2}\sqrt{\frac{1 + \beta^{-1} + \beta^{-1}_\mathrm{cr}}{3}}
    \label{eq:critical}
\end{eqnarray}
where $L_0$ is the outer scale of the turbulent cascade, $v_\mathrm{ph}$ is the phase velocity of waves, $T_4$ is gas temperature in units of $10^4\mathrm{K}$, and $0.15$ is a fitted coefficient found by \cite{2022Bustard} using simulation results. 

This diffusion coefficient is too low compared to $\kappa_\mathrm{MW}$. If $\kappa_\mathrm{crit}$ were the actual diffusion coefficient in the Milky Way, then the grammage would be increased to $X \sim 500 \,\mathrm{g}\,\mathrm{cm}^{-2}$. Even though $\kappa_\mathrm{crit}$ is physically unrealistic for an average value in the Milky Way, it is the other diffusion coefficient of interest because it leads to rapid cosmic ray energization \citep{2022Bustard}. 

In addition to a base simulation without cosmic rays (\texttt{No CR}) we run simulations with constant diffusion coefficients at $\kappa_\mathrm{MW}=3 \times 10^{28} \, \mathrm{cm}^2 \, \mathrm{s}^{-1}$ and $\kappa_\mathrm{crit}=5.82 \times 10^{25} \, \mathrm{cm}^2 \, \mathrm{s}^{-1}$, labelled \texttt{Milky Way} and \texttt{Critical} respectively (see Table \ref{tab:sims}). Finally, we run a simulation, labelled  \texttt{Two Zone}, which uses our two zone model detailed in Section \ref{sec:methods:decouple}. So we can more easily compare with the other simulations, we set the warm gas cosmic ray diffusion coefficient ($\kappa_w$ in Equation \ref{eq:decouple}) to the critical value $\kappa_\mathrm{crit}$ in Equation \ref{eq:critical}, and the cold gas cosmic ray diffusion coefficient ($\kappa_c$ in Equation \ref{eq:decouple}) to the Milky Way value $\kappa_\mathrm{MW}$.

\begin{figure*}[tbp]
    \centering
    \includegraphics[width=\textwidth]{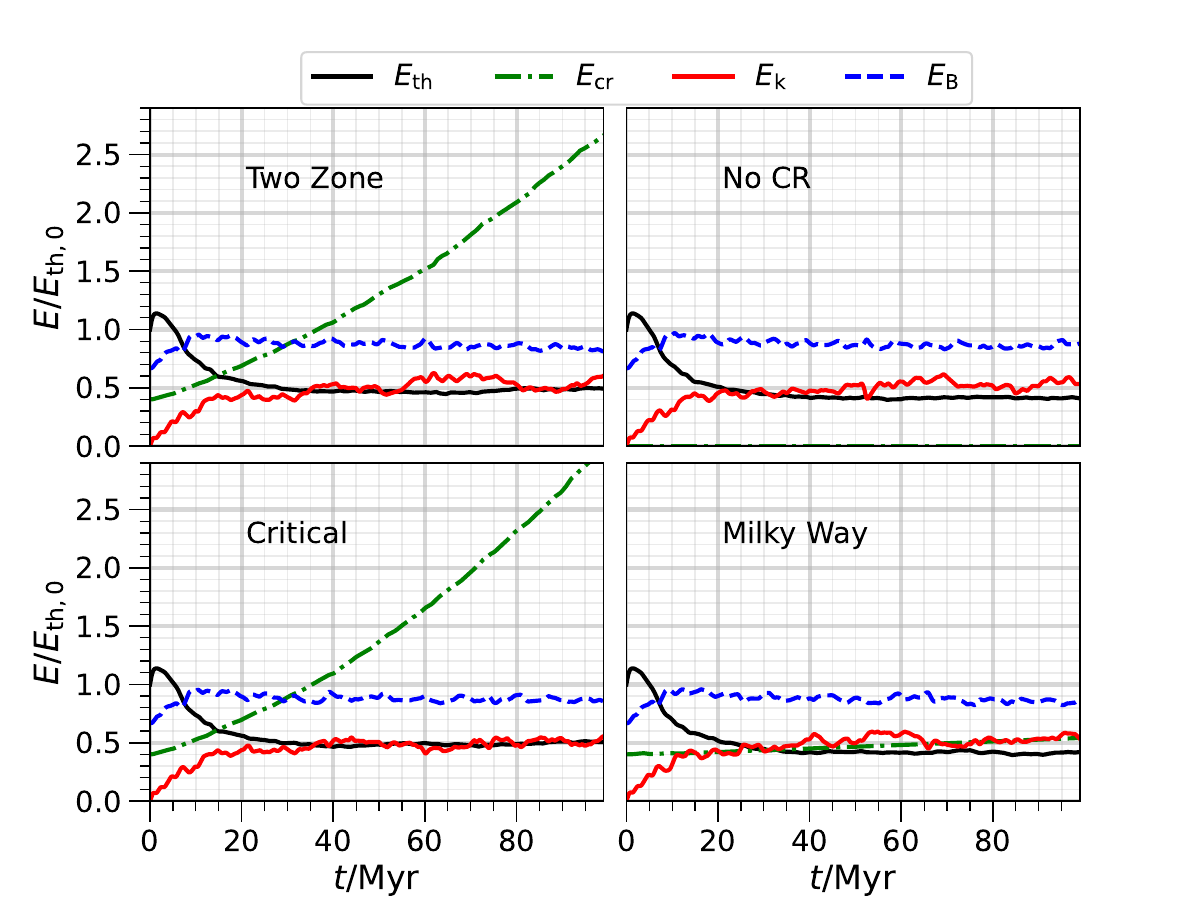}
    \caption{Energy evolution in each simulation. Solid black lines show the total thermal energy, dash-dotted green lines show the total cosmic ray energy, solid red lines show total kinetic energy, and dashed blue lines show total magnetic energy. All energies are normalized by the initial thermal energy  $(E_{\mathrm{th},0} = 3.9\cdot 10^{49}\mathrm{erg})$ in the simulation. While qualitatively similar, each run has unique characteristics. For example, the \texttt{Two Zone} and \texttt{Critical} simulations both have large increases in cosmic ray energy, but the cosmic ray energy in the \texttt{Two Zone} simulations grows at a slower rate (see Table \ref{tab:EradEturb}). }
    \label{fig:energy_evolution}
\end{figure*}

\begin{figure*}[btp]
    \centering
    \includegraphics[width=\textwidth]{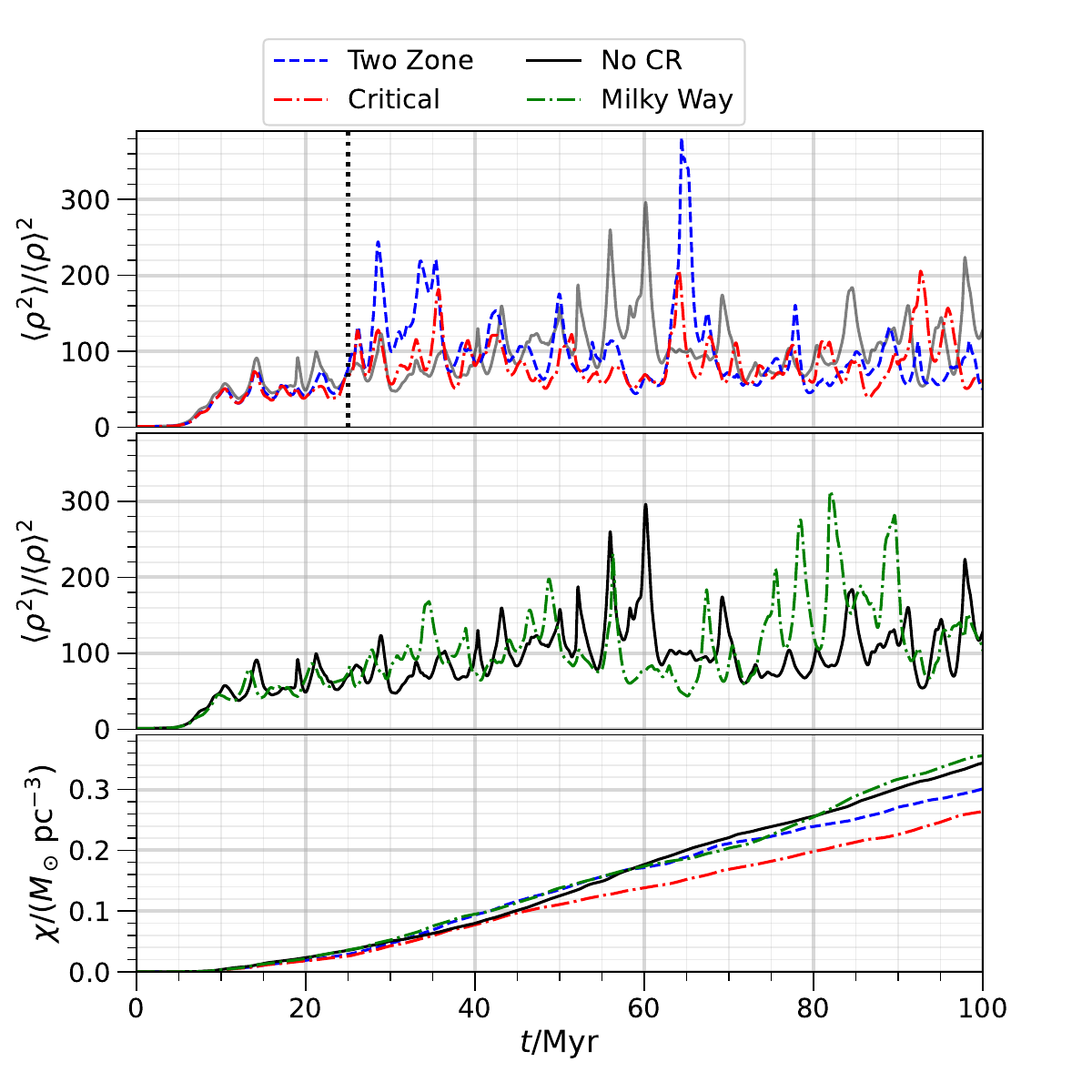}
    \caption{\textit{Top two plots:} Gas density clumping factor (as defined in \cite{2019Commercon}) for each simulation. Solid black line shows the \texttt{No CR} simulation, dash-dotted green line shows the \texttt{Milky Way} simulation, dash-dotted red line shows the \texttt{Critical} simulation, and dashed blue line shows the \texttt{Two Zone} simulation. To make the plot clearer, we separate the \texttt{Critical} and \texttt{Two Zone} from the \texttt{Milky Way} and \texttt{No CR} runs. Initially, the runs track with each other almost exactly. The \texttt{Milky Way} run ends up slightly out of phase quickly, likely because the cosmic rays can rapidly react to any compression and escape along the magnetic field. The \texttt{Two Zone} and \texttt{Critical} runs track with each other almost exactly for $20\,\mathrm{Myr}$. Then, the \texttt{Two Zone} run develops enough cold gas such that the cosmic ray transport differs significantly and more clumps of cold gas form. The \texttt{Two Zone} simulation then reaches clumping factors similar to those in the \texttt{No CR} and \texttt{Milky Way} simulations.
    \textit{Bottom plot:} Average value of scalar dye in each simulation (with units of $M_\odot \mathrm{pc}^{-3}$, see Section \ref{sec:methods:dye}). The \texttt{Critical} run produces the least amount of scalar dye at all times. The \texttt{Two Zone} simulation tracks with the \texttt{Milky Way} simulation once it has developed cold gas, while the \texttt{Milky Way} and \texttt{No CR} simulations fluctuate near one another for the entire simulation run.  }
    \label{fig:clump_evolution}
\end{figure*}

\begin{figure*}[btp]
    \centering
    \includegraphics[width=\textwidth]{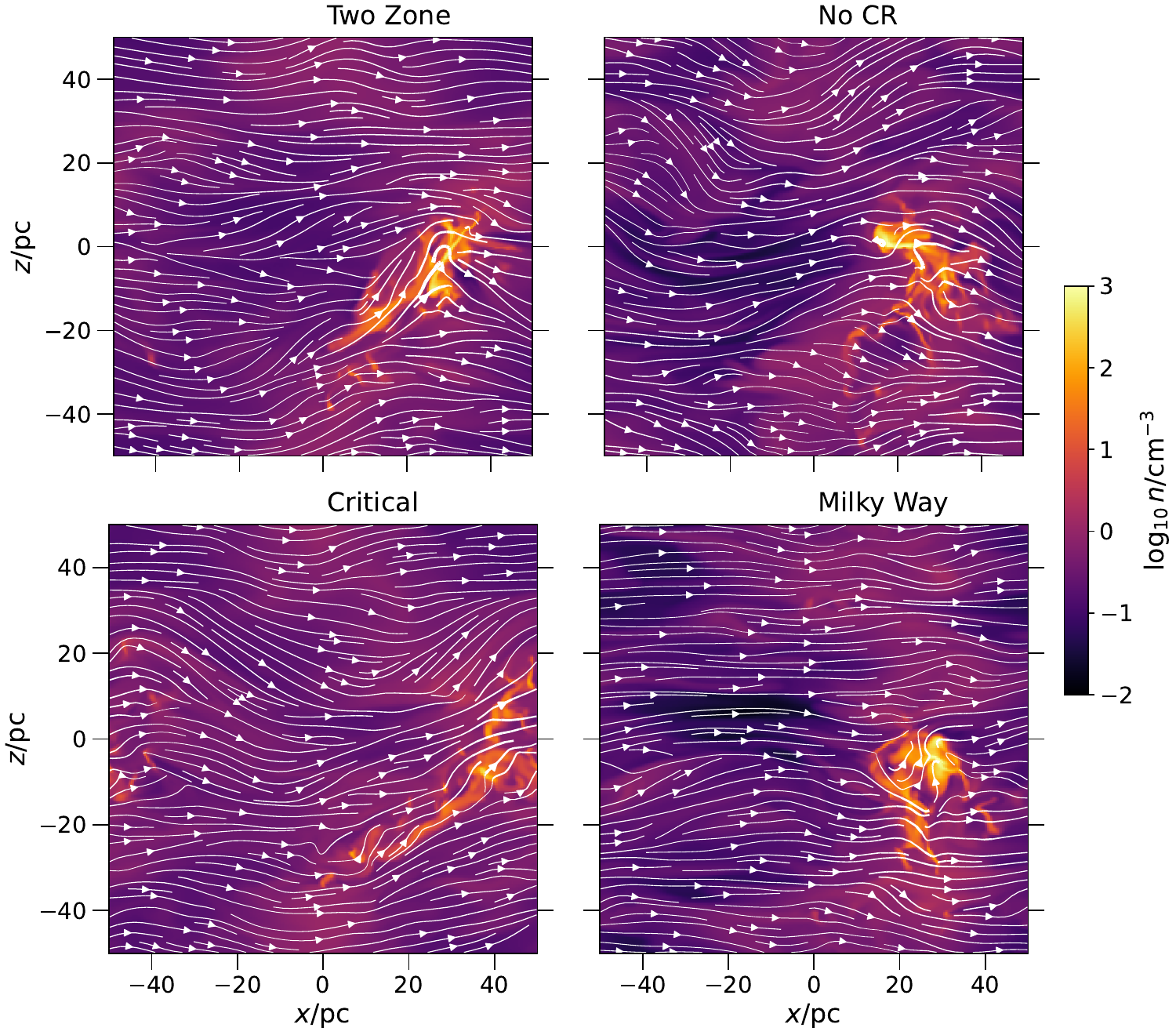}
    \caption{Slices from each simulation showing the gas number density. White lines show the magnetic field, and have a thickness proportional to the strength of the magnetic field (thicker line means stronger field). Each slice is at a time of $t=85 \,\mathrm{Myr}$ and at $y=0$. The density colormap is the same for each plot and is shown in the colorbar to the right.}
    \label{fig:clumpVisual}
\end{figure*}

\section{Results} \label{sec:results}
In this section we cover our simulations in depth. For a short summary of key takeaways, see Section \ref{sec:conclusions}. Each simulation is run for $100 \,\mathrm{Myr}$. That time frame allows the turbulent state to saturate and covers $\sim 8 $ sound crossing times $\tau_s = L_0/\sqrt{k_B T_i / m} \approx  13 \,\mathrm{Myr}$. When calculating median quantities in Sections \ref{sec:results:structure}, \ref{sec:results:diff}, \& \ref{sec:results:spectra} we set $25 \, \mathrm{Myr}$ to be the start of the saturated state and take the median over a time range of $[25,100]\,\mathrm{Myr}$ with data at each $1\,\mathrm{Myr}$. This saturated state after $t=25\,\mathrm{Myr}$ (shown as a dotted vertical line in the top plot of Figure \ref{fig:clump_evolution}) is apparent in Figures \ref{fig:energy_evolution} and \ref{fig:clump_evolution}. However, the saturated state is also highly intermittent. When studying that state we default to taking the median over time, instead of averaging, because it is more robust to large variations (e.g. the set of numbers $(1,1,11,1,1)$ has a median of $1$ but an average of $3$). Here, we show results from the $N=256$ simulations, but these results are also apparent in the $N=128$ simulations (see Appendix \ref{sec:appendix:LowRes}).

\subsection{Energy Evolution} \label{sec:results:energy}

\begin{table}[t]
    \centering
    \begin{tabular}{|l|r|r|}\tableline
        Simulation & $\dot{E}_\mathrm{rad} / \dot{E}_\mathrm{turb}$  &  $\dot{E}_\mathrm{cr}/  \dot{E}_\mathrm{turb}$\\\tableline
        \texttt{No CR} & $-1.3 \pm 0.3$ & NA \\\tableline
        \texttt{Milky Way} & $-1.3 \pm 0.5$ & $0.010 \pm 0.005$ \\\tableline
        \texttt{Critical} & $-1.1 \pm 0.3$ & $0.21 \pm 0.07$ \\\tableline
        \texttt{Two Zone} & $-1.1 \pm 0.5$ & $0.17 \pm 0.06$ \\\tableline
    \end{tabular}
    \caption{Median time derivatives of radiative energy and cosmic ray energy over the saturated regime from $t=25\,\mathrm{Myr}$ to $t=100\,\mathrm{Myr}$. The standard deviation during that time frame is shown after the $\pm$. By comparing the net cooling of the four simulations, it is apparent that having cosmic rays transported at the critical rate in warm gas siphons some energy, decreasing the energy lost to optically thin cooling.  The \texttt{Two Zone} and \texttt{Milky Way} simulations have the largest variation in radiative energy, likely because of their formation of dense structures (see Section \ref{sec:results:structure}). \label{tab:EradEturb}}
\end{table}

The evolution of thermal, cosmic ray, kinetic, and magnetic energy for each simulation is shown in Figure \ref{fig:energy_evolution}. The compressive driving initially heats the gas (increasing thermal energy) before some of the gas is pushed into a thermally unstable regime and collapses into cold clouds. This collapse leads to a loss of thermal energy as the gas cools via optically thin radiative cooling (see Section \ref{sec:methods:cooling}). 

The median radiative energy loss rate during the saturated regime is shown in Table \ref{tab:EradEturb}, normalized by the turbulent energy injection rate in Equation \ref{eq:driving}. The \texttt{No CR} and \texttt{Milky Way} simulations emit more radiation than the \texttt{Critical} and \texttt{Two Zone} simulations. 

Given the nearly constant loss of radiative energy (see Table \ref{tab:EradEturb}) along with the steady thermal, kinetic, and magnetic energy (see Figure \ref{fig:energy_evolution}), it is clear
that most of the energy injected is radiated away in the \texttt{No CR} and \texttt{Milky Way} simulations. The only difference in the \texttt{Critical} and \texttt{Two Zone} simulation is the steady increase of cosmic ray energy. Since the other energies reach a similar steady state as in the \texttt{No CR} simulation, there must be less radiation in the simulations with cosmic rays transported at the critical diffusion coefficient.

In Figure \ref{fig:energy_evolution}, the cosmic ray energy increases almost linearly in the three simulations which include cosmic rays, but the rate is dependent on the diffusion coefficient. In the \texttt{Critical} and \texttt{Two Zone} simulations, the increase in cosmic ray energy is faster than in the \texttt{Milky Way} simulation. In Table \ref{tab:EradEturb}, we show the median rate of change of cosmic ray energy over the saturated regime for each simulation. The rate of increase in the \texttt{Critical} and \texttt{Two Zone} simulations is approximately $20\times$ the rate of increase in the \texttt{Milky Way} simulation. This difference is the result of using the critical diffusion coefficient in the warm gas $\kappa_{w}=\kappa_\mathrm{crit}$. Because the warm gas is volume filling, $\kappa_w$ is the diffusion rate for a majority of the simulation's volume. With a significant volume near the critical value identified in \cite{2022Bustard} and \cite{2023Bustard}, there is cosmic ray energization (or reacceleration) which creates a larger energy sink. 

One striking feature of Figure \ref{fig:energy_evolution} is the nearly constant values of magnetic, kinetic, and thermal energy from $t=25 \,\mathrm{Myr}$ to $t=100 \,\mathrm{Myr}$. For each simulation, we do statistical analysis over that time frame to better understand the effect of including cosmic rays. A caveat to this is the increasing cosmic ray energy in the \texttt{Critical} and \texttt{Two Zone} simulations. However,the energy growth rate is consistent over the same time frame, and is a result of siphoning of energy from the rest of the ISM. The cosmic ray pressure gradients will be the important dynamical factor, and those can stay at the same order of magnitude while the total cosmic ray energy increases.

\subsection{Structure Formation}\label{sec:results:structure}

Next, we examine the formation of cold, dense gas in each simulation. \cite{2019Commercon} use a unitless clumping factor $\langle \rho^2 \rangle / \langle \rho \rangle^2$ to average over the volume of each of their simulations and quantify the appearance of dense structures. We calculate this clumping factor for each of our simulations, and show its evolution over time in Figure \ref{fig:clump_evolution}. The burstiness in time exhibited by the clumping factor highlights the necessity of frequent data dumps from the simulations. The lines in Figure \ref{fig:clump_evolution} are made up of data points from every $10^{-2}\,\mathrm{Myr}$, as we set the clumping factor to be calculated as part of \texttt{Athena++} history output.

The simulations track with each other early on, but eventually depart from each other in Figure \ref{fig:clump_evolution}. This initial consistency is because the simulations all use the same random seed to generate turbulence. Once they depart from each other, the \texttt{Two Zone} simulation consistently produces larger clumping factors than the \texttt{Critical} simulation, even reaching a higher peak value than the \texttt{No CR} and \texttt{Milky Way} simulations. This increase compared to the \texttt{Critical} simulation reflects the ability of the \texttt{Two Zone} simulation to create larger dense structures than the \texttt{Critical} simulation. If we label any of the increases in clumping factor as ``compressive episodes'', then it is clear those episodes occur at the same time in the \texttt{Critical} and \texttt{Two Zone} simulation. However, the adjustment of $\kappa_\parallel$ in cold gas allows higher density structures to form in the \texttt{Two Zone} simulation when compared to the \texttt{Critical} simulation.

\begin{figure}
    \centering
    \includegraphics[width=0.45\textwidth]{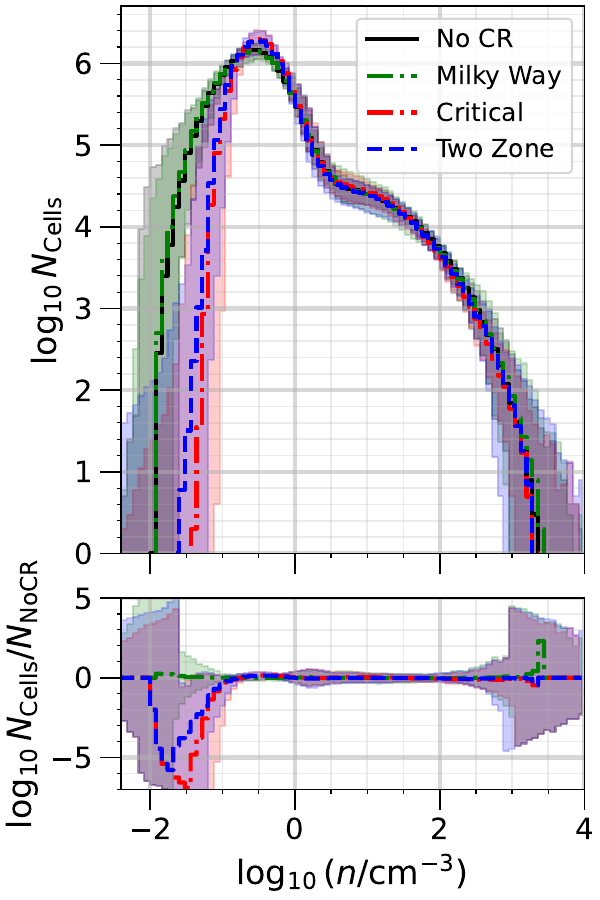}
    \caption{Number density histogram for each simulation. The lines show the median histogram over the entire period of saturation $(t=[25,100]\,\mathrm{Myr})$. The shaded regions show the full range of histogram values for each simulation. The \texttt{No Cr} and \texttt{Milky Way} simulation are nearly identical. The \texttt{Two Zone} and \texttt{Critical} simulations are also nearly identical, except the \texttt{Two Zone} simulation has more spread in the low density regime. The \texttt{Two Zone} and \texttt{Critical} simulations both produce less low density warm gas than the \texttt{Milky Way} and \texttt{No CR} simulations. }
    \label{fig:densHist}
\end{figure}

The \texttt{Milky Way} and \texttt{No CR} runs also depart from exact agreement, even ending up completely out of phase in their compressive episodes after $t > 70\,\mathrm{Myr}$. However, the variation during these runs is similar. The early deviation of compressive episodes is likely caused by the rapid diffusion of cosmic rays out of compressed regions. This diffusion along the magnetic field allows cosmic ray pressure gradients to form perpendicular to the magnetic field and affect the evolution of the gas density. However, in further analysis of the simulations (e.g. Figure \ref{fig:densHist}), we will see the \texttt{Milky Way} and \texttt{No CR} runs are very similar when averaged over time. 

Additionally, we show the evolution of the total amount of scalar dye $(\chi)$ (see Section \ref{sec:methods:dye}) in Figure \ref{fig:clump_evolution}. There is no decay or loss mechanism for $\chi$, so it tracks the total amount of cold dense gas produced over the course of each simulation. The \texttt{Critical} simulation produces the least amount of cold dense gas at all times, where as the \texttt{Two Zone}, \texttt{Milky Way}, and \texttt{No CR} simulations cross each other at different points. This crossing is caused by the intermittency of dense cloud formation, which is shown in the time evolution of the clumping factor in top plot of Figure \ref{fig:clump_evolution}. At the end of the simulation, the \texttt{Two Zone} simulation has produced over $20\%$ more cold dense gas than the \texttt{Critical} simulation, and the only difference between those simulations was the increased diffusion coefficient in cold gas. 

In Figure \ref{fig:clumpVisual}, we show slices of each simulation at the same positions $(y=0 \,\mathrm{pc})$ and time $(t=85\,\mathrm{Myr})$. These slices show the gas density with overlaid magnetic field as white streamlines. The thickness of the lines corresponds to the magnetic field strength at that point. The resolution is large enough to illustrate some of the smaller scale structure within the high density regions. The magnetic field is significantly distorted in the dense structures, where it is also the strongest. Overall, the magnetic field is predominantly in the $\hat{x}$ direction in each simulation.

The slices illustrate some of the key differences between the simulations. The \texttt{No CR} and \texttt{Milky Way} simulations reach lower densities in the volume filling warm gas (those slices have more black regions corresponding to $n\sim 10^{-2} \mathrm{cm}^{-3}$). The dense clump in the \texttt{Critical} simulation is not as distinct from the rest of the gas. The other simulations form more compact clumps along the magnetic field, with higher peak densities. This difference in structure likely results from the dense gas being less (or not) affected by cosmic ray pressure in the \texttt{Two Zone}, \texttt{Milky Way}, and \texttt{No CR} simulations.

\begin{figure}
    \centering
    \includegraphics[width=0.45\textwidth]{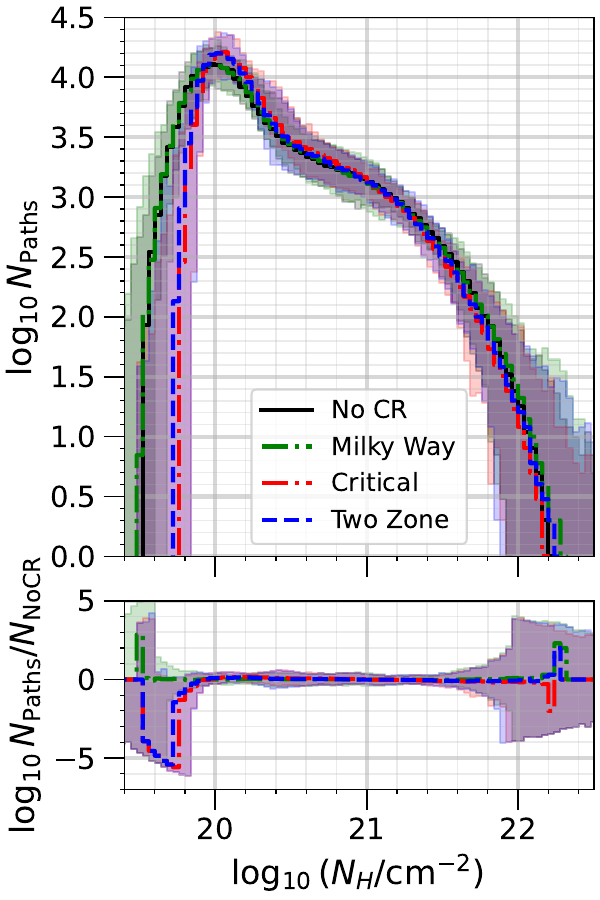}
    \caption{Column density histogram for each simulation. The line shows the median histogram over the entire period of saturation $(t=[25,100]\,\mathrm{Myr})$. The shaded regions show the full range of histogram values for each simulation. Each column density value is calculated by integrating along either the $\hat{x}$, $\hat{y}$ or $\hat{z}$ axis. This integration gives $256^2$ paths for each direction. The histogram is made from considering all $3\times 256^2$ paths we get from integrating along each axis individually.}
    \label{fig:columnHist}
\end{figure}

\begin{figure*}
    \centering
    \includegraphics[width=\textwidth]{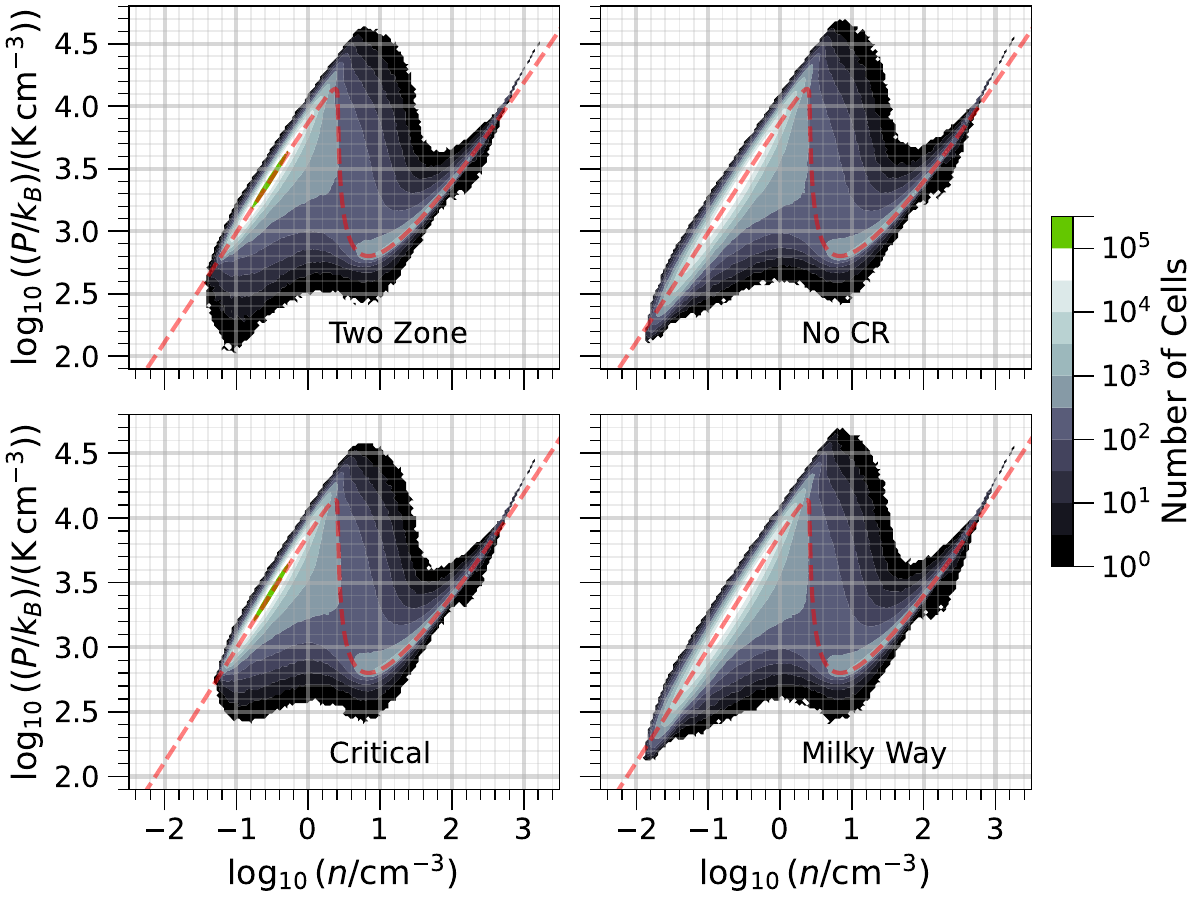}
    \caption{Pressure-Density phase diagrams for each simulation. The histograms illustrate the distribution of gas in phase space and along the thermal equilibrium curve $(\mathcal{L} = 0)$, which is shown as a dashed red line. Each diagram uses the same coloring for the contours, which is illustrated in the colorbar to the right of the plots. We take the median of the histograms over the saturated time frame, between $25\,\mathrm{Myr}$ and $100\,\mathrm{Myr}$. The \texttt{Two Zone} and \texttt{Critical} simulations have more warm gas concentrated above a number density of $10^{-1} \,\mathrm{cm}^{-3}$. Overall, the phase diagrams are similar across all the simulations. }
    \label{fig:phase}
\end{figure*}

\subsection{Gas Density \& Pressure Phase Space}\label{sec:results:phase}

Next we examine the phase space (gas pressure and number density) of each simulation. In Figure \ref{fig:densHist} we show the histogram of gas number density for each simulation.  Each shaded region corresponds to the full range of histograms for a given simulation, between $t=25\,\mathrm{Myr}$ and $t=100\,\mathrm{Myr}$ with $1\,\mathrm{Myr}$ time steps. Each line shows the median histogram over that time frame. Overall, the \texttt{No CR} and \texttt{Milky Way} simulations match, and the \texttt{Critical} and \texttt{Two Zone} simulations match. The \texttt{Two Zone} and \texttt{Critical} simulations have very little gas below $n=0.1\,\mathrm{cm}^{-3}$, compared to the other simulations. 

While the difference is not statistically significant, the \texttt{Critical} simulation has less high density gas, and more thermally unstable gas (near $n=10 \,\mathrm{cm}^{-3}$). This change likely results from the gas being more difficult to compress. With the smaller diffusion coefficient in the \texttt{Critical} simulation, the cosmic ray energy will not diffuse away fast enough during compressions. The cosmic ray pressure gradient then builds up and resists the compression. The compressibility of cosmic rays also limits the production of low density gas in the \texttt{Two Zone} simulation. However, because the diffusion coefficient changes and the cosmic ray pressure can rapidly diffuse along the magnetic field, the unstable gas in that simulation can collapse to the cold phase. This process leads to the \texttt{Two Zone} simulation's density histogram agreeing more with the \texttt{Milky Way} and \texttt{No CR} simulations in the higher density regime.

In Figure \ref{fig:columnHist} we show the distribution of column densities, integrated over one direction in each simulation. The distributions are, overall, similar to the number density histograms in Figure \ref{fig:densHist}. However, $N_H$ is an observable and we can now compare our distribution to realistic Milky Way column densities $N_H \sim 10^{21} \,\mathrm{cm}^{-2}$. Most integrated paths through our simulation do have a lower column density by an order of magnitude. These paths have a decreased column density because they either do not intersect a cold, dense cloud or cover enough distance. To match the Milky Way's column densities $(N_H =10^{21}\mathrm{cm}^{-2})$ for a majority of paths through our simulation, we would need to integrate across several realizations of our simulation volume. Then, more paths would intersect a cold cloud or pass through enough warm gas to reach observed column densities. This point should be kept in mind when evaluating the verisimilitude of the cosmic ray transport model in the \texttt{Two Zone} simulation. 

Next, we look at  pressure-density phase space. We produce phase space histograms for each $1\,\mathrm{Myr}$ of evolution in the saturated regime, and then we take the median of those histograms to produce a single plot for each simulation. Those median phase space distributions are shown in Figure \ref{fig:phase}. For each simulation, we also plot the thermal equilibrium curve as a dashed red line. The curve is determined by solving Equation \ref{eq:radiation} with $\mathcal{L} = 0$. 

\begin{table*}
    \centering
    \begin{tabular}{|l|ccc|ccc|} \tableline
        &  \multicolumn{3}{c|}{VFF (\%)} &\multicolumn{3}{c|}{MFF (\%)} \\\tableline
        Simulation  &  Cold & Warm & Unst. & Cold & Warm & Unst.   \\\tableline
        \texttt{No CR} & $ 1.6 \pm 0.2 $ & $92 \pm 1$ & $6\pm 1$ & $61 \pm 2$& $30\pm 1$ & $9 \pm 3$ \\\tableline
        \texttt{Milky Way} & $ 1.6 \pm 0.3 $ & $93 \pm 1$ & $6\pm 1$ & $60 \pm 2$& $31\pm 1$ & $10 \pm 2$ \\\tableline
        \texttt{Critical} & $ 1.6 \pm 0.2 $ & $93 \pm 1$ & $5\pm 1$ & $55 \pm 2$& $36\pm 1$ & $9 \pm 2$ \\\tableline 
        \texttt{Two Zone} & $ 1.5 \pm 0.2 $ & $93 \pm 1$ & $5\pm 1$ & $56 \pm 2$& $35\pm 1$ & $9 \pm 2$ \\\tableline 
    \end{tabular}
    \caption{Volume Filling Fraction (VFF) and Mass Filling Fraction (MFF) in each simulation, written as percentages. The quantities are calculated by averaging the MFF and VFF over the saturated time frame. We separate the phases by temperature: the cold phase is any gas below $200\,\mathrm{K}$, the unstable (Unst.) gas is anything with temperatures between the warm and cold gas, and the warm gas is any gas with temperature greater than $5500\,\mathrm{K}$. We also provide the standard deviation $(1\sigma)$ of the filling fractions over that $75\,\mathrm{Myr}$ of evolution. The only significant change due to cosmic rays and their transport is in the MFF of cold and warm gas. This change is related to the lower production of cold dense gas shown in Figure \ref{fig:clump_evolution}. \label{tab:ffs}}
\end{table*}

Each simulation's histogram is visually similar, with large populations of cold and warm gas in the thermally stable regimes of the equilibrium curve. There is more warm gas than cold gas in each simulation. We see the same cutoff of $n\sim 0.1\,\mathrm{cm}^{-3}$ for the \texttt{Two Zone} and \texttt{Critical} simulations which was apparent in the density histograms (Figure \ref{fig:densHist}). We also see the warm gas distribution in the \texttt{Two Zone} simulation extends to lower pressures at that density. This extension was also apparent at low densities in the density histograms, where the \texttt{Two Zone} simulations had slightly more low density gas than the \texttt{Critical} simulation. This extension is unique to the \texttt{Two Zone} simulation.

The turbulent driving is compressive, creating regions with increased density and pressure. This driving is more apparent in the \texttt{No CR} and \texttt{Milky Way} phase diagrams. The extension toward the top of the phase diagram in these simulations is larger because the cosmic ray energy is taking away less of the injected energy during a compression. In the \texttt{Two Zone} and \texttt{Critical} simulations, the warm gas has the critical diffusion coefficient, allowing the cosmic ray energy to serve as a sink for some of the injected turbulent energy, reducing the amount of compression which takes place. 

Finally, in Table \ref{tab:ffs}, we show the volume filling fractions (VFFs) and mass filling fractions (MFFs) of the different gas phases in each simulation. The listed quantities are percentages, and they are calculated by taking the median over the saturated time frame. The listed errors are the standard deviation during the $75\,\mathrm{Myr}$ saturated time frame. Overall, the simulations all have the same VFF for each phase. The only impact cosmic rays have is on the MFF of cold and warm gas. This decrease in cold gas MFF (and corresponding increase in warm gas MFF) is likely tied to the lower total amount of cold gas formation over the course of the simulation (see scalar dye evolution in Figure \ref{fig:clump_evolution}). The change in MFF with no equivalent adjustment of VFF suggests the median density in cold gas is lower in the \texttt{Critical} and \texttt{Two Zone} simulations, and similarly the median density in the warm gas is higher. This difference also shows up in the phase diagrams (see Figure \ref{fig:phase}) and the density histograms (see Figures \ref{fig:densHist} \& \ref{fig:columnHist}).

\begin{figure}
    \centering
    \includegraphics[width=0.45\textwidth]{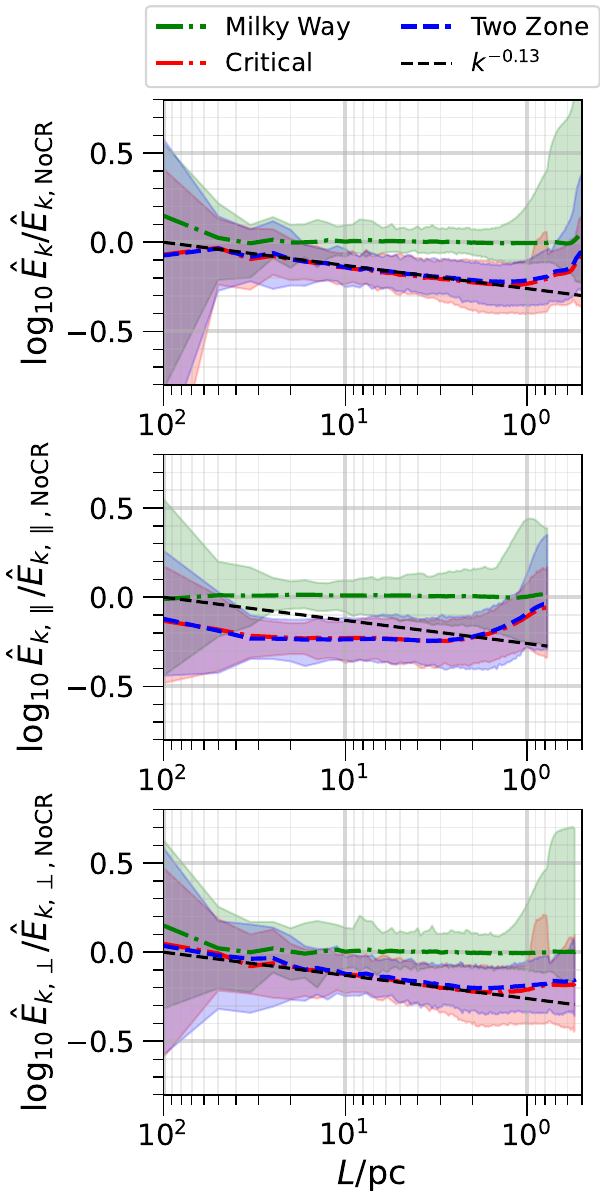}
    \caption{Kinetic energy per mass spectra from the \texttt{Milky Way}, \texttt{Critical}, and \texttt{Two Zone} simulations compared to the \texttt{No CR} simulation's spectrum. Lines show the median spectrum, whereas the shaded regions show the full variation of the spectrum over the saturated timeframe $(t=[25,100]\,\mathrm{Myr})$. \textit{Top plot:} The total kinetic energy per unit mass. \textit{Middle plot:} The kinetic energy per unit mass parallel to the mean magnetic field. \textit{Bottom plot:} The kinetic energy per unit mass perpendicular to the mean magnetic field. }
    \label{fig:kinSpec}
\end{figure}
\begin{figure}
    \centering
    \includegraphics[width=0.45\textwidth]{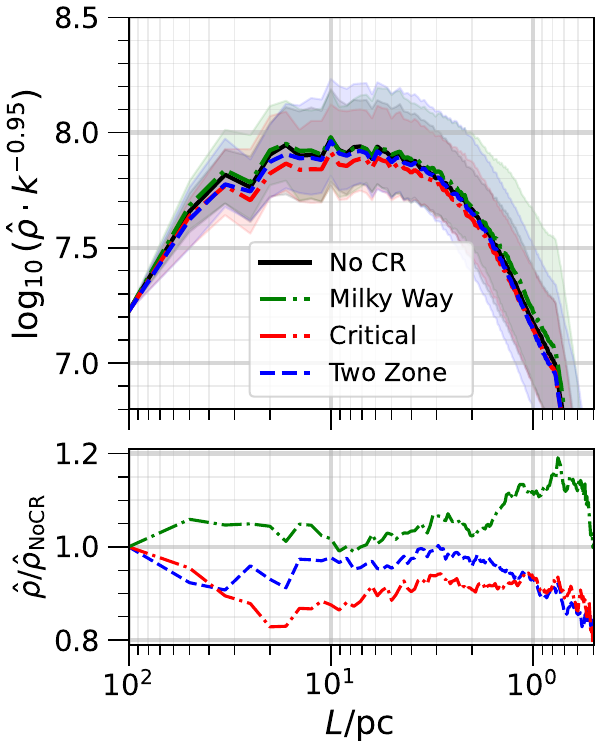}
    \caption{\textit{Top plot:} Fourier transform (spectrum) of gas density for each simulation, scaled by a power law of $k^{0.95}$. The lines are the median spectrum, whereas the shaded regions show the full variation over the saturated time frame. \textit{Bottom plot:} The ratio of other simulations' median spectrum to the \texttt{No CR} simulation spectrum. }
    \label{fig:densSpec}
\end{figure}

\begin{table}
    \centering
    \begin{tabular}{|l|c|}\tableline
        Simulation & Kinetic Energy Slope ($\alpha$ in $k^{-\alpha}$) \\\tableline
        \texttt{No CR} &  $1.79 \pm 0.07$ \\\tableline
        \texttt{Milky Way} & $1.78 \pm 0.06$ \\\tableline
        \texttt{Critical} &  $1.91 \pm 0.08$\\ \tableline
        \texttt{Two Zone} &  $1.91 \pm 0.07$ \\ \tableline
    \end{tabular}
    \caption{Median inertial range spectral index for each simulation with $1\sigma$ variations. The dependence on $\kappa_w$ is clear, as simulations \texttt{Critical} and \texttt{Two Zone} have a steeper slope and are the simulations with cosmic ray energy diffusion at the critical rate $\kappa_\mathrm{crit}$ through the warm gas. Steepening of the spectrum as a result of cosmic ray transport matches results from \cite{2023Bustard} even though we have multiple gas phases. Additionally, the slope of our \texttt{No CR} run is similar to the slope of the equivalent run in \cite{2023Bustard}. \label{tab:inertial}}
\end{table}

\subsection{Power Spectra} \label{sec:results:spectra}

To study the effect of the cosmic ray energy on the saturated turbulence, we examine how the power spectrum of kinetic energy varies between each simulation.  In Figure \ref{fig:kinSpec}, we plot the total kinetic energy per mass $(v^2/2)$ spectrum, the spectrum parallel to the mean magnetic field (meaning the $\hat{x}$ direction, so $v_x^2/2$), and the spectrum perpendicular to the mean magnetic field. The solid lines correspond to the median spectrum over the saturated time frame and the shaded regions show the full variation. We divide the spectra from the \texttt{Milky Way}, \texttt{Critical}, and \texttt{Two Zone} simulations by the median spectrum of the \texttt{No CR} simulation. 

Figure \ref{fig:kinSpec} illustrates that the \texttt{Two Zone} and \texttt{Critical} simulations have a significantly different turbulent cascade than the \texttt{No CR} simulation. The cosmic rays have little to no effect in the \texttt{Milky Way} simulation, where the cosmic rays are not well coupled to the outer eddy scale. Additionally, the middle and bottom panels of Figure \ref{fig:kinSpec} illustrate that the well coupled cosmic rays in the \texttt{Two Zone} and \texttt{Critical} simulations have an anisotropic impact on the turbulent cascade. While the parallel kinetic energy at all scales is reduced, the perpendicular kinetic energy experiences an actual change in slope. The steepening of the slope is similar to some results in \cite{2023Bustard}, even though our simulations have multiple gas phases. The change in slope is approximately $k^{-0.13}$ in the \texttt{Critical} and \texttt{Two Zone} simulations. For completeness, we calculate the original slope of each spectrum's inertial range, and these are listed in Table \ref{tab:inertial}. For each $1\,\mathrm{Myr}$ time dump, we fit a power law $k^{-\alpha}$ over the length scales $L\in [ 5,20]\,\mathrm{pc}$. We then take the median of the power law index $\alpha$ over the saturated time frame to get the median slope values and their $1\sigma$ variation.

In the top plot of Figure \ref{fig:densSpec} we show the Fourier transform of the gas density in each simulation. In the bottom plot of Figure \ref{fig:densSpec} we show the ratio of the median Fourier transform for each simulation, relative to the median of the \texttt{No CR} simulation. Overall, the simulations produce a similar characteristic density spectrum, with a power law $\sim k^{-0.95}$ in the inertial range ($L\in [ 5,20]\,\mathrm{pc}$, see Figure \ref{fig:kinSpec}). The \texttt{Critical} simulation has less dense gas (lower Fourier amplitudes) at all scales. However, as the bottom of Figure \ref{fig:densSpec} shows, the variations in the inertial range are on the scale of $10\%$ when comparing the simulations. From the top plot, it is clear that the variation over time for each simulation (the shaded regions) is even larger than the variation between simulations.

\begin{figure}
    \centering
    \includegraphics[width=0.45\textwidth]{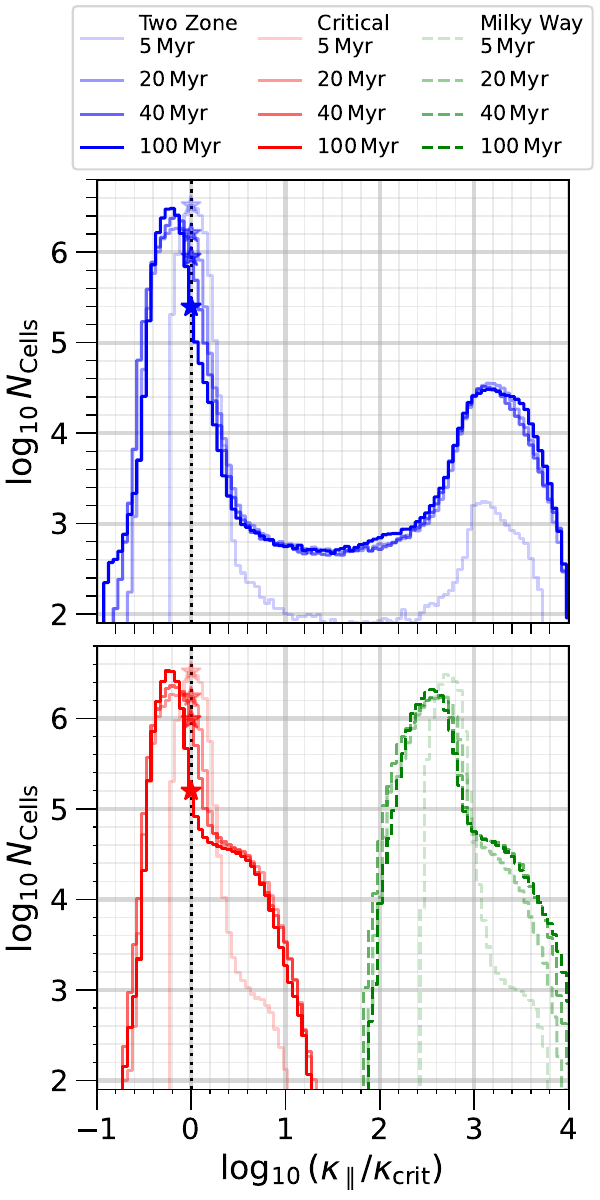}
    \caption{Histograms of $\kappa_\parallel/\kappa_\mathrm{crit}$ for the \texttt{Two Zone} (top plot, blue lines), \texttt{Critical} (bottom plot, red lines), and \texttt{Milky Way} (bottom plot, green lines) simulations. At each cell, we calculate $\kappa_\mathrm{crit}$ from Equation \ref{eq:critical} using the cell's pressure and density values. We then take the actual diffusion coefficient in that cell and divide by the local critical coefficient. The histograms are shown at several times, and stars mark where $\kappa_\parallel = \kappa_\mathrm{crit}$. }
    \label{fig:KappaRatio}
\end{figure}

\begin{figure}
    \centering
    \includegraphics[width=0.49\textwidth]{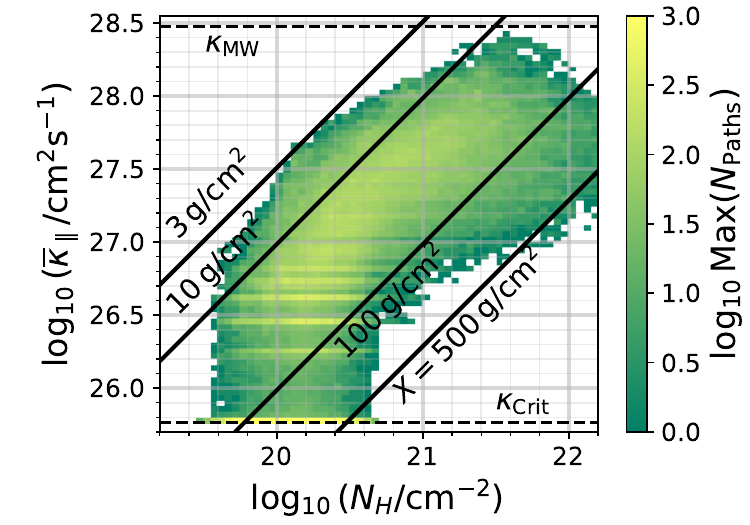}
    \caption{
    Maximum distribution of average cosmic ray diffusion coefficient and gas column density along the mean magnetic field (in the $\hat{x}$ direction) for the \texttt{Two Zone} simulation. For each $1\,\mathrm{Myr}$ time dump in the saturated state $(t=[25,100]\,\mathrm{Myr})$, we calculate the histogram in $(N_\mathrm{H}, \overline{\kappa}_\parallel)$ space. To produce this figure, we take the maximum of each bin across time. The resulting distribution, shown here, presents the full range of diffusion coefficients and column densities produced in the simulation. In solid black lines, we plot contours of grammage (see Equation \ref{eqn:grammage}), illustrating that at certain times in the simulation we get realistic values (order unity in units of $\mathrm{g} \cdot \mathrm{cm}^{-2}$) of the grammage for some paths along the mean magnetic field. We also show the asymptotic values of the diffusion coefficient, $\kappa_\mathrm{crit} = 5.82 \cdot 10^{25} \mathrm{cm}^2 \mathrm{s}^{-1}$ and $\kappa_\mathrm{MW}  = 3 \cdot 10^{28} \mathrm{cm}^2 \mathrm{s}^{-1}$, as horizontal dashed lines.}
    \label{fig:KappaHist}
\end{figure}

\subsection{Diffusion Coefficient Statistics} \label{sec:results:diff}

In Figure \ref{fig:KappaRatio}, we show histograms of the ratio of the cosmic ray energy diffusion coefficient $\kappa_\parallel$ (calculated from Equation \ref{eq:decouple}) to the local critical diffusion coefficient $\kappa_\mathrm{crit}$ (calculated at each cell using Equation \ref{eq:critical}) for the \texttt{Two Zone}, \texttt{Critical}, and \texttt{Milky Way} simulations. The blue lines in the upper plot are from the \texttt{Two Zone} simulation. The red lines in the lower plot are from the \texttt{Critical} simulation. The green lines in the lower plot are from the \texttt{Milky Way} simulation. Stars mark the value of the histogram at $\kappa_\parallel=\kappa_\mathrm{crit}$.

The stars move down with time in both the \texttt{Two Zone} and \texttt{Critical} simulations. This decrease in the amount of gas with cosmic ray transport at the critical rate is the result of increasing cosmic ray energy in the simulations (see Figure \ref{fig:energy_evolution}). As the cosmic ray energy increases, the critical diffusion coefficient $\kappa_\mathrm{crit}$ increases. This increase leaves a majority of the gas with a diffusion coefficient below the critical value. 

The distributions of the \texttt{Milky Way} and \texttt{Critical} simulations primarily reflect the gas temperature distribution, because they have constant diffusion coefficient $\kappa_\parallel$ but a critical diffusion coefficient $\kappa_\mathrm{crit}$ dependent on gas temperature, cosmic ray pressure, and magnetic pressure (see Equation \ref{eq:critical}). The \texttt{Two Zone} simulation looks similar, except the cold gas component has been shifted to the right because of its larger diffusion coefficient value.

In Figure \ref{fig:KappaHist}, we show the relationship between average cosmic ray diffusion coefficient $\overline{\kappa}_\parallel$ and column density $N_H$ for paths along the mean magnetic field in the \texttt{Two Zone} simulation. For each $1\mathrm{Myr}$ time dump in the saturated time frame $(t=[25,100]\,\mathrm{Myr})$, we integrate along the $\hat{x}$ direction, parallel to the mean magnetic field, to find the column density and average diffusion coefficient. For each time dump, this gives $256^2=65536$ paths. Then, we take the maximum across all times in each bin of $(N_H,\overline{\kappa}_\parallel)$ space and plot those in Figure \ref{fig:KappaHist}. Note that the \texttt{Milky Way} and \texttt{Critical} simulations would appear as single horizontal lines in this figure, located at their particular value of the diffusion coefficient (either $\kappa_\mathrm{MW}$ or $\kappa_\mathrm{crit}$, shown as dashed lines in Figure \ref{fig:KappaHist}).

A majority of the paths travel only through warm gas and keep the original critical diffusion coefficient value, showing up as a thin region of yellow bins in the bottom of Figure \ref{fig:KappaHist}. However, for the paths which pass through some cold gas, the average diffusion coefficient changes significantly. This change also affects the grammage, which can be calculated using Equation \ref{eqn:grammage}. Contours of the grammage are shown as black lines in Figure \ref{fig:KappaHist}, and some of the paths end up with realistic values of the grammage $X$ (order unity in units of $\mathrm{g} \cdot \mathrm{cm}^{-2}$ according to observable estimates, \citealt{2001Jones,2019Evoli}). 

Figure \ref{fig:KappaHist} is a lower limit on the distribution of average diffusion coefficient $\overline{\kappa}_\parallel$. Using mass-weighted average or actual path integration along the magnetic field, the paths would pass through more cold gas because there is some bending in the field lines (see Figure \ref{fig:clumpVisual}). This change in paths would cause more gas at high diffusivity because more magnetic field lines pass through the cold gas (where the field strength is larger). Additionally, our lines of integration do not pass through as much of the ISM as cosmic rays in the Milky Way (see column density histogram in Figure \ref{fig:columnHist}). If every path passed through a cold cloud, then there would be many more paths with a realistic mean diffusion coefficient $\overline{\kappa}_\parallel$.

Figure \ref{fig:KappaHist} helps illustrate the key takeaway from our use of a temperature dependent diffusion coefficient in the \texttt{Two Zone} simulation. The temperature dependence allows cosmic rays to significantly affect the turbulent energy cascade in warm gas (see Figure \ref{fig:kinSpec}). However, this was previously only possible if the diffusion coefficient was too low compared to Milky Way observations (e.g. the results of \citealt{2019Commercon} and \citealt{2023Bustard}). By having the diffusion coefficient increase in cold gas, we keep the effect on the cascade while the average diffusion coefficient, and resulting grammage, is closer to observational estimates.

\section{Discussion} \label{sec:disc}
This section roughly follows the order of results presented in Section \ref{sec:results}. For our final conclusions and summary, see Section \ref{sec:conclusions}.

Figure \ref{fig:energy_evolution} together with Table \ref{tab:EradEturb} show that when cosmic rays are weakly coupled or not present at all, the system reaches a steady state with turbulent energy input balanced by compression and radiative cooling. When strongly coupled cosmic rays are added to the mix, as in the \texttt{Critical} and \texttt{Two Zone} simulations, some turbulent energy input goes into compressive heating of cosmic rays. In that case, less of the turbulent energy input is lost to radiative cooling. 

However, there is rarefaction as well as compression, and one might ask why the cosmic rays gain energy instead of returning it back to the flow during rarefaction. A full answer to this question is beyond the scope of the present paper and we plan to address it in future work. However, we provide a partial answer by considering how cosmic ray diffusion affects small amplitude acoustic waves. 

Consider the linearized cosmic ray pressure equation for waves propagating along the ambient magnetic field:
\begin{equation}
    \pdv{\delta P_c}{t}=-\gamma_c P_c \pdv{\delta v}{x}+\kappa_\parallel \pdv[2]{\delta P_c}{x}.
    \label{eq:lineng}
\end{equation}
Assume the solution is a plane wave with frequency $\omega$ and wavenumber $k$. Cosmic rays do work on the gas at the rate $\dot{E} = -\delta v\partial\delta P_c/\partial x$ (note $E$ refers to energy density, not energy). Using the linearized equation to determine $\delta P_c$ in terms of $\delta v$ allows us to calculate the energy change
\begin{equation}
    \dot{E} = -\gamma_c P_c \vert k\delta v \vert^2
    \frac{\kappa_\parallel k^2}{\omega^2+ \kappa_\parallel^2 k^4}.
    \label{eq:deltaE}
\end{equation}
Equation (\ref{eq:deltaE}) shows that cosmic rays steadily extract energy from the wave, much as a gas which conducts heat extracts energy from sound waves. The effect is maximized at $\omega/k =\kappa_\parallel k$. If we identify $\omega/k$ with the phase velocity and $k$ with $L^{-1}$ we recover the argument used to estimate the critical diffusivity $\kappa_\mathrm{crit}$ (see Equation \ref{eq:critical}) in \cite{2022Bustard}.

This effect has a counterpart in the kinetic theory description of cosmic ray propagation in the limit of short mean free path \citep{1975Skilling}. It is akin to first order Fermi acceleration because the cosmic rays are scattering off converging fluctuations. Note that our simulations do not include actual shock acceleration. We would need strong shocks as well as an explicit term (e.g. shock acceleration term in \cite{2019Dubois}) which more accurately models the acceleration process on subgrid scales.

We reiterate that this linear perturbation argument only considers interaction with a simple longitudinal sound wave. This derivation does not account for the interaction of the cosmic rays with a turbulent cascade. The apparent dependence of the total energy of the thermal gas on the diffusion coefficient $(\kappa_\parallel^{-1})$ would change when considering the entire cascade. Specifically, the compression in multiphase turbulence is much more frequent than in adiabatic turbulence \citep{instability}. We expect that the $\nabla \cdot v$ term (first term of right hand side of Equation \ref{eq:lineng}) will have a stronger effect in terms of energization of cosmic rays. 

Returning to the effect of cosmic rays on an entire turbulent cascade, we see that during the formation of dense gas, the cosmic rays can diffuse independently of the gas, quickly redistributing energy from compressed regions. Those dense regions can collapse further because any cosmic ray pressure gradients have been decreased or removed. This process is clearest in the evolution of the gas density clumping factor and scalar dye $\chi$ in Figure \ref{fig:clump_evolution}. We see the \texttt{Two Zone}, \texttt{Milky Way}, and \texttt{No CR} simulations have large spikes and variation in clumping as opposed to the evolution in the \texttt{Critical} simulation where cosmic rays provide significant pressure support against the formation of cool clouds. While cool clouds still form in the \texttt{Critical} simulation, there are fewer of them and they do not reach as high densities, which is clear from the \texttt{Critical} simulation's smaller production of scalar dye $\chi$. This process agrees with the results presented in \cite{2019Commercon}, which showed that decreasing the diffusion coefficient leads to smoother gas distributions and fewer cold clouds. 

Moving on to cosmic ray transport through this multiphase ISM, the \texttt{Two Zone} simulation allowed us to examine how a temperature dependent diffusion coefficient would affect the grammage cosmic rays pass through on their way through the ISM. Observations of the primary-to-secondary ratio give stringent requirements on the grammage $X$ that $\mathrm{GeV}$ cosmic rays will experience on average \citep{2019Evoli,2020Evoli}. In our \texttt{Two Zone} simulation, the cosmic rays mainly experience a grammage set by the volume filling warm gas. However, Figure \ref{fig:KappaHist} shows that some paths will have physically realistic grammage values when only the cold gas has a realistic diffusion coefficient. Additionally, most paths through the simulation do not produce a realistic column density (see Figure \ref{fig:columnHist}). Performing the integration over multiple realizations of the simulation box would produce more sightlines with higher column densities. Those higher column density sightlines could also have a higher average diffusion coefficient $\overline{\kappa}_\parallel$ if they intersect cold clouds.

Therefore, for Milky Way column densities, the diffusion coefficient in the cold gas could become the dominant contributor to the mean cosmic ray diffusion coefficient. This conclusion adds to the importance of cosmic ray transport modelling which accounts for changes in temperature, ionization, or magnetic field strength. For example, in the realistic case of the diffusion coefficient in the warm gas and cold gas being $\kappa_\parallel \sim \kappa_\mathrm{MW}$, the diffusion coefficient in hot and warm ionized gas could be a much lower value (possibly near $\kappa_\mathrm{crit}$) without changing the average cosmic ray diffusion coefficient calculated from the observed primary-to-secondary ratio. 

While not a physically significant result, we find that the numerical stability of the \texttt{Two Zone} simulation depends on how we calculated the diffusion coefficient. Applying the switching function in logarithmic space leads to a smoother transition in regions with a large temperature gradient. This may be useful for other studies which model inhomogeneous diffusion without the added computational expense of calculating $\kappa_\parallel$ \textit{ab initio} and on the fly from gas parameters.

We show the cosmic ray diffusion coefficient has a significant impact on the turbulent cascade in Section \ref{sec:results:spectra}. This results from a simple relationship in the original CR+MHD equations. A rapid flattening and smoothing of cosmic ray pressure, caused by a large diffusion coefficient, minimizes the effect of cosmic rays because there is less time for the pressure gradient $\gradient P_c$ to change the motion of gas (Equation \ref{eqn:mom}). A smaller diffusion coefficient means large peaks in the cosmic ray pressure can exist longer and create large pressure gradients which adjust the gas flow.

The steepening of the kinetic energy spectrum by a factor of $k^{-0.13}$ in Figure \ref{fig:kinSpec} is clearly the result of the critical diffusion coefficient in the warm gas because the same results appear for simulations \texttt{Critical} and \texttt{Two Zone}. The reduction in parallel kinetic energy at all scales is because of the decreased diffusion parallel to the magnetic field,  i.e. the $\kappa_\parallel \nabla^2 E_c$ term contained in the $\div{\vb{F_c}}$ term of Equation \ref{eqn:cre}. The lower diffusion coefficient means there is enough time for the turbulent flows to drive more energization before the cosmic rays diffuse away.

The steepening in spectral slope for the perpendicular (and total) kinetic energy is more difficult to physically explain. If the $k^{-1.79}$ dependence of the total kinetic energy spectrum in simulation \texttt{No CR} is the result of our turbulent driving being too weak to develop a \citeauthor{K41} $(k^{-5/3})$ or \citeauthor{1964Iroshnikov}-\citeauthor{1965Kraichnan} $(k^{-3/2})$ spectrum, then the further adjustment towards Burgers turbulence $(k^{-2})$ in the \texttt{Two Zone} and \texttt{Critical} simulations suggests transport at the critical diffusion coefficient may increase the frequency of `shocks' in the direction perpendicular to the magnetic field. Although, we see little to no formation or propagation of shocks. Instead, it is possible the sharp density discontinuities between the cold and warm phases have an effect on turbulent transport similar to that of shocks in Burgers turbulence (instantaneous transfer of energy from large to small scales rather than a true cascade).

A secondary point from the examination of the kinetic energy spectra is that the slope of the \texttt{No CR} run is similar to the slope found in \cite{2023Bustard}, despite our inclusion of optically thin radiative cooling. This similarity is likely because motions on inertial scales $(20\,\mathrm{pc}>L> 5 \,\mathrm{pc})$ are mostly determined by the warm gas in our simulations. This warm gas follows an equation of state similar to the simple adiabatic law in \cite{2017NJPh...19f5003K} \& \cite{2023Bustard}. To see the differences created by radiative cooling, we would need to extend the inertial scale below $1 \,\mathrm{pc}$. This extension requires increasing the resolution of our simulations. However, an increase in resolution will make our treatment of diffusive transport of $E_c$ inaccurate (see Section \ref{sec:methods:CRMHD}). This limitation highlights the importance of future simulations which evolve the cosmic ray distribution function and/or actual particle trajectories (e.g. using methods like MHD-PIC, \citealt{Athena_MHDPIC}).

\rh{For clarity, we would like to quickly highlight the key differences and similarities between our study and previous studies of \cite{2019Commercon}, \cite{2019Dubois}, \cite{2022Bustard}, and \cite{2023Bustard}. First, no other study has quantified and shown the anisotropic nature of cosmic ray feedback on a multiphase ISM turbulent energy cascade (see Figure \ref{fig:kinSpec}). Second, our \texttt{Two Zone} simulation, which incorporates a temperature dependent diffusion coefficient (motivated by the global simulations of \citealt{2018Farber}), is unique; it has no counterpart in previous studies. Thirdly, the idea that cosmic rays can take energy which would otherwise be radiated during thermal instability (see Table \ref{tab:EradEturb}) has not been shown previously. 

There are clear similarities between our work and previous works. Our \texttt{No CR}, \texttt{Milky Way}, and \texttt{Critical} simulations are directly comparable to some simulations in \cite{2019Commercon}. Our turbulent cascade results are similar to those in \cite{2023Bustard}. However, our simulations are run for a longer simulation time and on larger scales than \cite{2019Commercon}, and \cite{2023Bustard} did not incorporate a multiphase ISM or thermal instability. Therefore, even where our results are similar, our work is different such that it adds to the robustness of some key ideas and conclusions in \cite{2019Commercon} and \cite{2023Bustard}. Overall, we view our work as an extension of these previous works, while also illustrating new results and possibilities.}

Prior to concluding, it is important to note that a significant caveat to our results is our simulations are longer than the cosmic ray confinement time in the galaxy. Therefore, most of the cosmic ray energy should escape the simulation box. Over that time frame, cosmic ray losses to hadronic and Coulomb interactions could also be significant on the time scales we simulate. Using a hadronic loss term integrated over cosmic ray energy from \cite{2007Enslin,2021Bustard}, the timescale for losing cosmic ray energy to hadronic interactions with the thermal gas is 
\begin{equation}
    \tau_\mathrm{Hadron} = \frac{E_\mathrm{CR}}{\dot{E}_\mathrm{CR}} = c n \bar{\sigma}_{pp} K_p = 66\,\mathrm{Myr}.
    \label{eqn:hadron}
\end{equation}
We decided to not include the effect of hadronic losses, or a realistic confinement time, to focus on the coupling between the thermal gas and cosmic ray fluid already inherent in Equations \ref{eqn:cont} to \ref{eqn:crflux}. Since we did not include these effects, it is important to note the significant cosmic ray energy in the \texttt{Critical} and \texttt{Two Zone} simulations would likely escape the galaxy, rather than being kept in the ISM, because of the small confinement time of cosmic rays. If the confinement time were longer than the hadronic loss time, then the energy would end up in photons emitted during hadronic interactions.

An additional caveat is our focus on diffusive and advective transport of cosmic rays. We do not include streaming transport, nor its heating effect. Some simulations in \cite{2023Bustard} examine the impact of streaming on the turbulent cascade. However, for the high densities inside the galactic disk $(n>1\mathrm{cm}^{-3})$, it is expected that cosmic rays are transported diffusively \citep{2021Armillotta,2022Armillotta,2023Thomas,2024Armilotta}. Therefore, focusing on diffusive transport in this study is a reasonable reduction in scope.

\section{Conclusions} \label{sec:conclusions}
We presented simulations of a bi-stable ISM including the effects of cosmic ray feedback for three different cosmic ray diffusion coefficient models, along with a simulation with no cosmic ray energy. We used a long baseline for statistical analysis of the saturated turbulent state. We showed that cosmic rays can decrease the amount of energy radiated away via optically thin cooling. This energy siphoning is mediated by a process similar to first order Fermi acceleration in regions with a non-zero divergence of the velocity field. We illustrated the impacts of a temperature dependent (as a proxy for ionization dependence) cosmic ray diffusion coefficient on the formation of cold dense gas and cosmic ray grammage. We re-examined the adjustment of the turbulent energy cascade by cosmic rays detailed in \cite{2022Bustard,2023Bustard} and find the cosmic rays have an anisotropic effect on the cascade.

From these simulations, our analysis, and our discussion, the key conclusions are:
\begin{enumerate}
    \item Cosmic ray transport at the critical rate allows cosmic rays to remove energy from collapsing gas via first order Fermi acceleration, decreasing the total radiated energy (see Table \ref{tab:EradEturb}).
    \item Increasing the diffusion coefficient only within cold gas allows $\sim 20\%$ more cold, dense gas to form over time than in a simulation with a constantly low diffusion coefficient (see Figure \ref{fig:clump_evolution}).
    \item Even when the volume filling warm gas has a small diffusion coefficient some paths through our simulation box have an average grammage near Milky Way values $(\gtrsim 1 \,\mathrm{g}\,\mathrm{cm}^{-2}$, see Figure \ref{fig:KappaHist}). Given that our column densities are low (see Figure \ref{fig:columnHist}), we actually underestimate the number of paths through the ISM with a larger diffusion coefficient. This underestimation also applies to the number of paths with realistic grammage.
    \item Cosmic ray energy transported via diffusion at the critical rate identified in \cite{2023Bustard} decreases kinetic energy parallel to the mean magnetic field at all scales by nearly a factor of $2$, while also changing the spectral slope of kinetic energy perpendicular to the mean magnetic field by a factor of $k^{-0.13}$ (see Figure \ref{fig:kinSpec}).
\end{enumerate}

\section*{Acknowledgments}
We would like to thank Ryan Farber and Chad Bustard for sharing their expertise and advice during the course of this work. We also acknowledge Peng Oh for extensive discussions during the final stage of the paper. We thank Michael Halfmoon for granting additional NERSC time needed for this project. Finally, we would like to thank the referee for many useful comments and suggestions which significantly improved this work.

RH acknowledges funding from NASA FINESST grant 80NSSC22K1749 and NSF grant AST-2007323 during the course of this work. 
Research presented in this article was supported by the LDRD program of LANL with project \# 20220107DR (KWH) \& 20220700PRD1 (KHY), and a U.S. DOE Fusion Energy Science project. 
This research used resources provided by the LANL Institutional Computing Program (y23\_filaments), which is supported by the DOE NNSA Contract No. 89233218CNA000001. This research also used resources of NERSC with award numbers FES-ERCAP-m4239 (PI: KHY) and m4364 (PI: KWH).

\vspace{5mm}

\software{Athena++ \citep{2020Stone,2018Jiang}, MatPlotLib \citep{2007Matplotlib}, NumPy \citep{2011NumPy,2020NumPy}, AstroPy \citep{2013AstroPy,2018Astropy}}



\begin{figure*}
    \centering
    \includegraphics[width=1.0\textwidth]{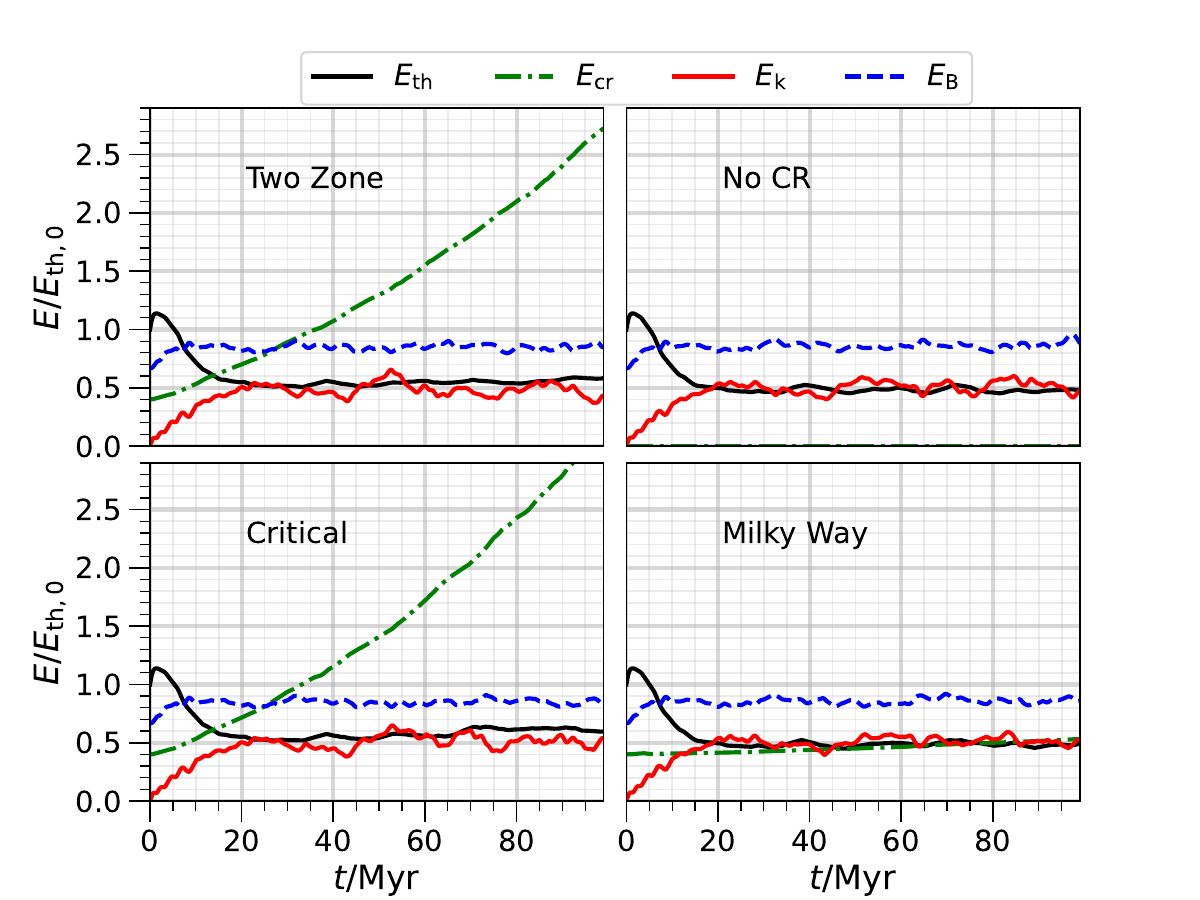}
    \caption{Same as Figure \ref{fig:energy_evolution}, but for the lower resolution $N=128$ simulations. }
    \label{fig:energy_evolution_128}
\end{figure*}

\begin{figure}
    \centering
    \includegraphics[width=0.45\textwidth]{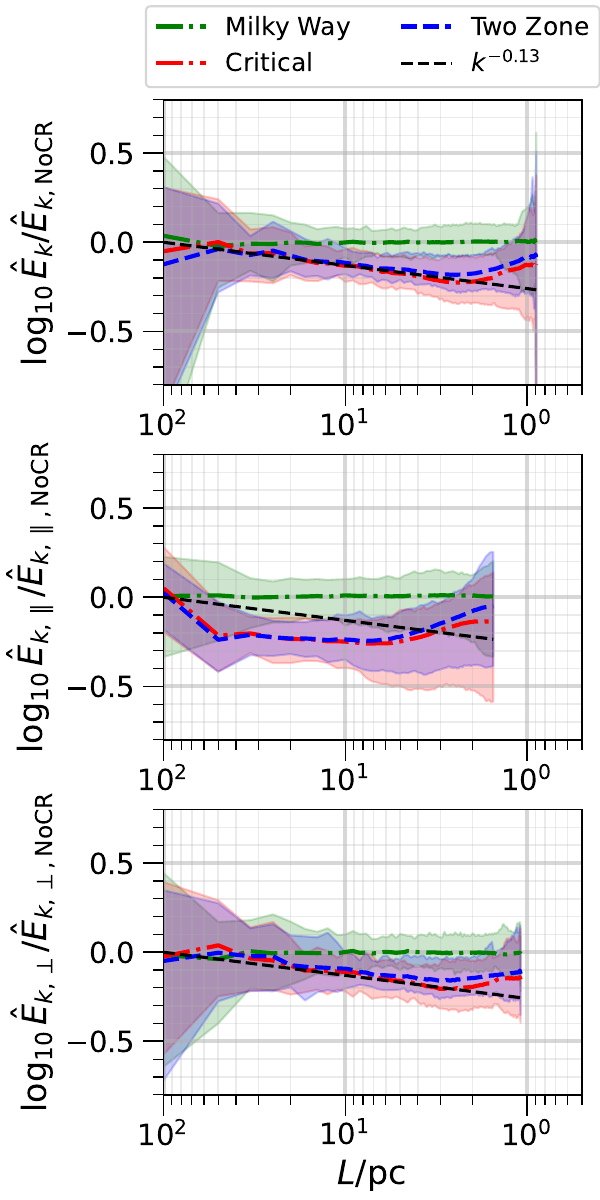}
    \caption{Same as Figure \ref{fig:kinSpec}, except for the lower resolution $N=128$ simulations.}
    \label{fig:kinSpec128}
\end{figure}

\appendix
\section{Lower Resolution Results} \label{sec:appendix:LowRes}
We performed lower resolution simulations $(N=128, \Delta x \sim 0.78 \,\mathrm{pc})$ with the same setup as the simulations listed in Table \ref{tab:sims}. The results exhibit little to no difference from the higher resolution simulations $(N=256, \Delta x \sim 0.39 \,\mathrm{pc})$, illustrating the robustness of our conclusions. 

Figure \ref{fig:energy_evolution_128} shows the evolution of energy in the $N=128$ simulations. The time evolution of each energy is similar to the $N=256$ simulations, with the \texttt{Two Zone} and \texttt{Critical} simulations having significant cosmic ray energization.

Figure \ref{fig:kinSpec128} shows the kinetic energy cascade, and its parallel and perpendicular components. Even in the $N=128$ simulations with a smaller inertial range, the change in slope is the same for the overall cascade and for the parallel and perpendicular components.

\bibliographystyle{aasjournal}
\bibliography{refs}


\end{document}